%% file: CoLOD.tex
\documentclass[acmtog]{acmart}
\usepackage{wrapfig}
\acmSubmissionID{349}

\setcopyright{acmlicensed}
\acmJournal{TOG}
\acmYear{2024} 
\acmVolume{43} 
\acmNumber{6} 
\acmArticle{193} 
\acmMonth{12}
\acmDOI{10.1145/3687905}

\input{src/macros}

\begin{document}

\title{Architectural Co-LOD Generation}

\author{Runze Zhang}
\email{oliverzrz.cyber@gmail.com}
\orcid{0009-0008-6304-623X}
\affiliation{%
  \institution{Shenzhen University}
  \country{China}}

\author{Shanshan Pan}
\email{psshappystar@gmail.com}
\orcid{0009-0008-8259-5234}
\affiliation{
  \institution{Shenzhen University}
  \country{China}}

\author{Chenlei Lv}
\email{lvchenlei@gmail.com}
\orcid{0000-0002-8203-3118}
\affiliation{
  \institution{Shenzhen University}
  \country{China}}

\author{Minglun Gong}
\email{minglun@uoguelph.ca}
\orcid{0000-0001-5820-5381}
\affiliation{
  \institution{University of Guelph}
  \country{Canada}}

\author{Hui Huang}
\email{hhzhiyan@gmail.com}
\authornote{Corresponding author: Hui Huang (hhzhiyan@gmail.com)}
\orcid{0000-0003-3212-0544}
\affiliation{
  \department{College of Computer Science \& Software Engineering}
  \institution{Shenzhen University}
  \country{China}
}

\renewcommand{\shortauthors}{R. Zhang, S. Pan, C. Lv, M. Gong, and H, Huang}

\begin{abstract}

Managing the level-of-detail (LOD) in architectural models is crucial yet challenging, particularly for effective representation and visualization of buildings. Traditional approaches often fail to deliver controllable detail alongside semantic consistency, especially when dealing with noisy and inconsistent inputs. We address these limitations with \emph{Co-LOD}, a new approach specifically designed for effective LOD management in architectural modeling. Co-LOD employs shape co-analysis to standardize geometric structures across multiple buildings, facilitating the progressive and consistent generation of LODs. This method allows for precise detailing in both individual models and model collections, ensuring semantic integrity. Extensive experiments demonstrate that Co-LOD effectively applies accurate LOD across a variety of architectural inputs, consistently delivering superior detail and quality in LOD representations.

\end{abstract}



\begin{teaserfigure}
  \centering
  \includegraphics[width=\linewidth]{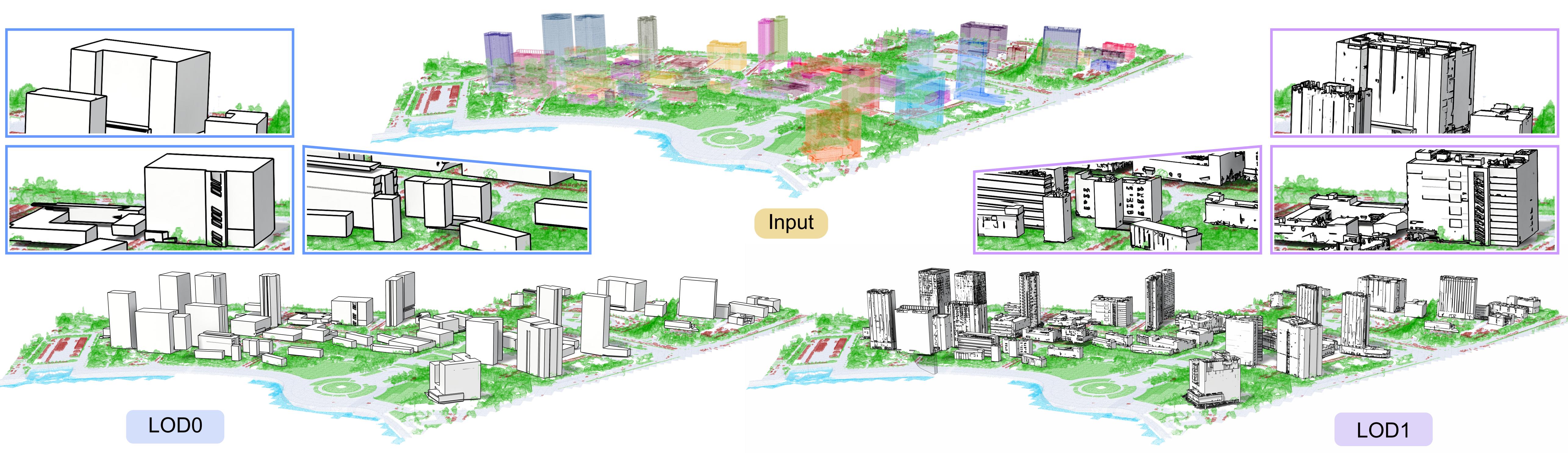}
  \caption{Co-LOD generation for architectural models from raw real-world point clouds.}
  \label{fig:teaser}
\end{teaserfigure}

\maketitle

\input{src/01-intro}
\input{src/02-rw}
\input{src/03-method}

\input{src/04-results}

\input{src/05-future}

\begin{acks}
This work was supported in parts by NSFC (U2001206, 62161146005, U21B2023), Guangdong Basic and Applied Basic Research Foundation (2023B1515120026, 2023A1515110292), DEGP Innovation Team (2022KCXTD025), Shenzhen Science and Technology Program (KQTD 20210811090044003, RCJC20200714114435012), NSERC (293127), and Scientific Development Funds from Shenzhen University.
\end{acks}

\bibliographystyle{ACM-Reference-Format}
\bibliography{CoLOD_ref}
\end{document}

%% file: src/macros.tex
\usepackage{color}
\usepackage{xcolor}
\usepackage{graphicx}
\usepackage{booktabs}
\usepackage{amsmath, amsthm, amsfonts, amssymb}
\usepackage{mathrsfs}
\usepackage{textcomp}
\usepackage{epstopdf}
\usepackage{multirow}
\usepackage{wrapfig}
\usepackage{subfig}
\usepackage{booktabs} 
\usepackage[ruled,linesnumbered]{algorithm2e} 
\usepackage{algorithmicx}  
\usepackage{algpseudocode}
\usepackage{ragged2e}
\usepackage[normalem]{ulem}
\usepackage{cleveref}
\usepackage{bm}

\usepackage{enumitem}

\SetAlFnt{\small}
\SetAlCapFnt{\small}
\SetAlCapNameFnt{\small}
\SetAlCapHSkip{0pt}

\newcommand{\eg}{\textit{e.g.}}

\definecolor{green}{rgb}{0, 0.5, 0}
\definecolor{orange}{rgb}{1, 0.5, 0.0}
\definecolor{red}{rgb}{1.0, 0.0, 0.0}
\definecolor{teal}{rgb}{0.0, 0.4, 0.4}
\definecolor{purple}{rgb}{0.65,0,0.65}
\definecolor{saffron}{rgb}{0.95,0.75,0.2}
\definecolor{lightblue}{rgb}{0.0,0.5,1}
\definecolor{brown}{rgb}{0.5, 0.16, 0.16}
\definecolor{brickred}{rgb}{.6, .2 .1}
\definecolor{coral}{rgb}{1,0.45,0.33}
\definecolor{newcolor}{rgb}{.8,.349,.1}

%% file: src/01-intro.tex
\section{Introduction}
\label{sec:intro}

In the realm of architectural model reconstruction, the creation of structural details at various levels is pivotal for a range of applications such as urban planning, autonomous driving, virtual reality, and digital entertainment. As the pioneering work, CityGML~\cite{cityGML} outlines that a 3D digital building can be depicted through multiple levels-of-detail (LODs), with each level providing progressively more detailed structures. This forms the basis of LOD generation with two significant challenges. Firstly, the intricate and diverse geometric structures of architectural models make it difficult to robustly control the generation of LODs. Secondly, maintaining semantic consistency within the same LOD level across different building models is a demanding task.

Existing approaches focusing on the LOD generation of building models have been attempting to address the aforementioned two challenges~\cite{low-poly,LOD2015,KSR}. To ensure semantic consistency across different models, they use predefined rules to control LOD generation. However, these rules often falter when dealing with complex geometric structures and non-standard inputs. To overcome the limitations of external rules, some methods utilize visual error control and geometric constraints to guide the LOD generation. The aim of these methods is to achieve a more universally applicable and robust reconstruction process that also maintains semantic consistency. Despite these efforts, consistently producing uniform LODs from diverse and intricate inputs remains challenging. The success of maintaining semantic consistency heavily relies on the quality of the input architectural models. At present, there is a lack of an effective mechanism for controlling LOD generation that can simultaneously meet the demands for robustness and semantic consistency.

In response to these challenges, we introduce Co-LOD, a novel approach that utilizes a co-analysis strategy. Co-LOD operates under the assumption that architectural models contain semantically meaningful details at different hierarchical levels, which are discernible through patterns of repetitiveness and correlation. By conducting a co-analysis of these repetitive and autocorrelated geometric structures across different input models, we can more precisely control the different LOD layers. Co-LOD harnesses the structural association relationships inherent in architectural models to develop correlation rules for LODs, thereby achieving semantic consistency. 

The process encompasses two primary stages: initially, raw inputs (\eg point clouds or triangle meshes) are segmented into structural segments to identify the main structure and categorize geometric details at varying levels. Subsequently, we formulate an energy function for these segments, which guides the LOD generation process while ensuring semantic consistency. When applied to a collection of building models, Co-LOD enhances the robustness against individual sample variations, offering an effective way to manage LOD generation. It strikes a balanced chord between robustness and semantic consistency, inheriting the advantages of structured reconstruction with concise and clear mesh results. Co-LOD is pioneering in its use of co-analysis to implement precise LOD controlling, an example of which can be seen in Fig.~\ref{fig:teaser}. The key contributions of this approach are summarized as follows:

\begin{itemize}
\item \textbf{Innovative Co-Analysis Approach.} Co-LOD represents a novel approach in integrating co-analysis for Level-of-Detail (LOD) generation. This initiative is the first to employ a building set-based cross-analysis specifically tailored for LOD generation tasks. Our method opens new research avenues, especially in achieving semantic consistency in the representation and optimization of architectural models.

\item \textbf{Structural Segments Generation.} Our methodology features a meticulously designed segmentation process, capable of handling a wide array of architectural inputs and producing uniform structural segments. This approach excels in handling complex geometric structures and minimizes dependency on specific parameters, thus significantly enhancing the robustness and adaptability for processing raw inputs.

\item \textbf{Joint Structural Analysis.} To augment our co-analysis, we have developed a comprehensive segment-based quantitative analysis method, tailored for architectural models. It establishes rules based on the hierarchy and similarity of structural details, which supports precise LOD control across LOD layers. This method effectively balances maintaining semantic consistency with ensuring robustness in LOD generation.
\end{itemize}

%% file: src/02-rw.tex
\section{Related Work}\label{sec:rw}

\noindent\textbf{Low-Poly Reconstruction.}
To achieve structural model from architectural point clouds, some researchers develop low-poly reconstruction methods that are used to generate initial LOD information.  \citet{mehra2009abstraction} utilized characteristic curves and contours to establish building blocks for low-poly representation. \citet{Co-abstract} implemented consistent abstraction of man-made models based on subvolumes. \citet{lafarge2013surface} proposed a structure-preserving
approach to detect planar components and construct polygonal mesh. \citet{li2016manhattan} presented a fully automatic approach for reconstructing urban scene from point samples. It implemented plane hypothesis and generated candidate boxes to approximate geometry structures of buildings. With the similar technical route, they improved the framework~\cite{nan2017polyfit} for reconstructing lightweight polygonal surfaces from point clouds. \citet{kelly2017bigsur} proposed a high-quality structural modeling method for city blocks, which formulates a binary integer program to produce semantic models
with associated surface elements. \citet{KSR} designed an efficient shape assembling mechanism to reconstruct watertight polygonal meshes from point clouds. It is capable of accurately representing piecewise planar structures while also approximating freeform objects from incomplete geometric structures. \citet{fang2018planar} proposed a structural method to detect planar shapes from 3D data. Then, they designed an hybrid approach~\cite{fang2020connect} to successively connects and slices planes for polygonal mesh reconstruction. Such solutions focus on individual model reconstruction without semantic consistency and precise LOD control.

\noindent\textbf{LOD Generation.}
A series of detailed re-factoring methods are proposed to generate LODs, which can be concluded into two groups: geometric-based and human observation-based. Methods of first group implement LOD generation based on geometric feature analysis. Some of them~\cite{Pms,QEM,FPS,RobustLowPoly,out_of_core,IDS,low-poly} utilized edge collapse strategy to control LOD with geometric error optimization. Inspired by marching cubes~\cite{Mcube}, some solutions~\cite{neuralLOD,neuralLOD_compress} generated LOD according to implicit surface estimation. Such methods can handle most 3D shape for LOD generation. However, they lack semantic perception abilities that may produce structural defects or inconsistencies in architectural models. Another group utilize human observations to improve ability of semantic-aware in LOD generation. \citet{LOD2015} learned geometric attributes and semantic rules to extract facades, roofs, and superstructures.  
Based on CityGML's~\cite{cityGML} requirements, such elements are allocated to suitable levels for LOD generation.
\citet{ArrangementNet} classified architectural point clouds into floor, wall, ceiling, and window components with similar purpose. Such methods rely on manually designed rules that take some instability factors for individual building models.

\begin{figure*}[t]
	\centering
	\includegraphics[width=1\textwidth]{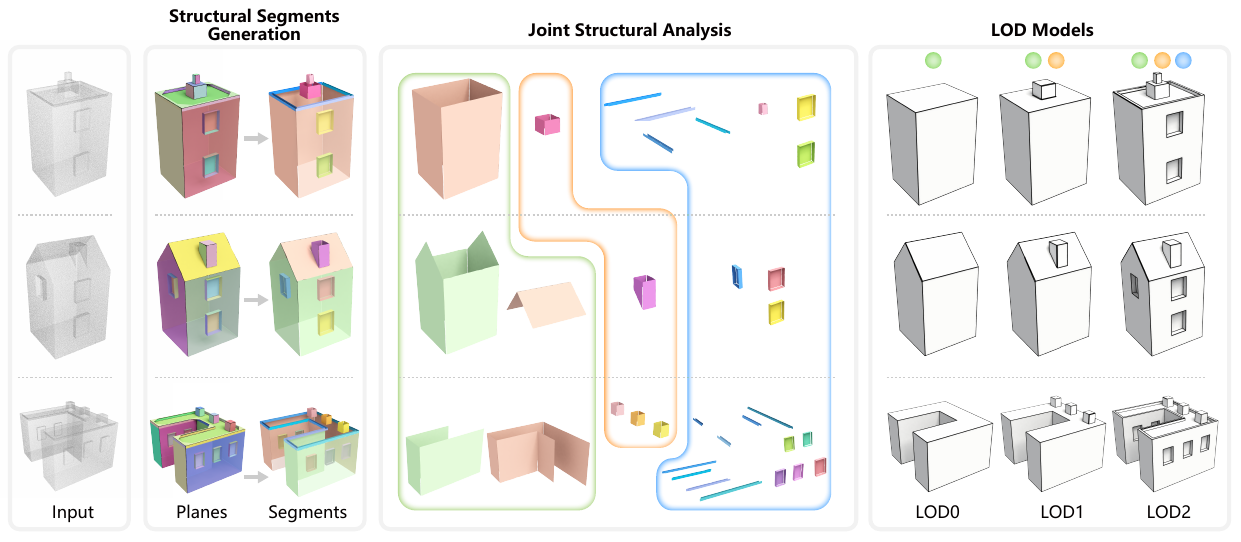}
	\caption{Overview of Co-LOD workflow, which contains two main modules. The first module involves detecting primary planes and aggregating them to form structural segments from raw inputs. The second module quantitatively measures the similarity of these segments and clusters them to define LOD layers, a process akin to joint structural analysis. Together, these modules enable Co-LOD to effectively perform co-analysis for controlled LOD generation with semantic consistency.}
	\label{fig:overview}
\end{figure*}

\noindent\textbf{Structural Segmentation.} Another important technical route is to utilize segmented components to guide LOD reconstruction. Initial methods for architectural segmentation utilized shallow pipelines incorporating hand-crafted point descriptors and rules~\cite{build_seg_1,build_seg_5,build_seg_6}. 
To improve the accuracy for repetitive structural segmentation, \citet{build_seg_3} formulated a weighted minimum set to implement segment optimization. \citet{build_seg_2} leveraged symmetry for architectural structural segmentation. \citet{build_seg_4} proposed a user-assisted segmentation method to further reduce ambiguous structures. Recently, \citet{BuildingNet} introduced a BuildingNet dataset that is used to support learning-based schemes. It is useful to define accurate segmentation or LOD component. However, such methods don't consider the relationships between different segmentations which limit the performance for accurate LOD controlling and semantic analysis. In non-architectural segmentation domains, some approaches~\cite{Co-1,Co-2,Co-3,Co-4,Co-5,Co-6} employed co-analysis to improve the accuracy of segmenting components. They initially employed mesh segmentation technique~\cite{Random_cuts,Fuzzy_cuts,SDF_seg} 
to over-segment the input model, followed by feature extraction and clustering to achieve a consistent structural segmentation. 
Our approach draws inspiration from this scheme and proposes the prior work to implement co-analysis for architectural LOD generation.

%% file: src/03-method.tex
\section{Methodology}

\noindent\textbf{Overview.} Co-LOD is designed to facilitate LOD generation for architectural models with controllable structural accuracy and semantic consistency. Specifically, Co-LOD takes a group of architectural models as input and generates LOD representations for these models that adhere to the following criteria:
\begin{itemize}
    \item [i)] Each input architectural model corresponds to an entity within each layer;   
    \item [ii)] Architectural models at the same layer exhibit LOD-based semantic consistency;   
    \item [iii)] Structures become progressively enriched with the increase in LOD;   
    \item [iv)] LODs are represented by watertight polygonal meshes.   
\end{itemize}
Essentially, Co-LOD achieves LOD generation that meets the specified criteria through three core steps, as illustrated in Fig.~\ref{fig:overview}. The first stage involves structural segmentation to identify basic components. This is followed by a joint structural analysis that conducts a co-analysis on these segments, organizing them into LOD layers with semantic consistency. Finally, LODs represented by polygonal meshes are extracted from these layered segments. The following sections will delve into the implementation details.

\subsection{Structural Segments Generation} \label{Segments Generation}

The importance of structural segmentation in LOD generation is well-recognized. Existing methods, however, often rely on low-level primitives for LOD generation, limiting their capacity to generate models at different levels-of-detail. To overcome this limitation, we propose a novel segmentation scheme aimed at grouping interrelated planes to obtain meaningful structural segments. It begins with the extraction of primary planes from a raw point cloud or mesh, which aligns with the initial steps of traditional methods~\cite{KSR,nan2017polyfit}. The scheme then progresses by using the detected planes to form structural segments, serving as the foundation for subsequent LOD control. 
\begin{wrapfigure}{r}{0.25\textwidth}
    \vspace{-0.5cm}
    \hspace{-0.5cm}
	\includegraphics[width=0.25\textwidth]{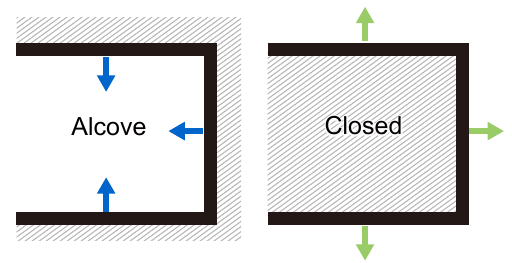}
    \vspace{-0.5cm}
\end{wrapfigure}
A challenge in performing structural segmentation on architectural models is the presence of two distinct types of structures: alcove and closed (see right). As outlined in Algorithm~\ref{algo1}, we address this challenge by initially inverting planes smaller than $A_\epsilon$ to convert alcoves into closed structures for extraction, and then utilizing the remaining planes for closed structure extraction. These generated segments effectively facilitate subsequent LOD-based analysis.

\begin{wrapfigure}{r}{0.13\textwidth}
	\hspace{-0.6cm}
	\centering
	\includegraphics[width=0.13\textwidth]{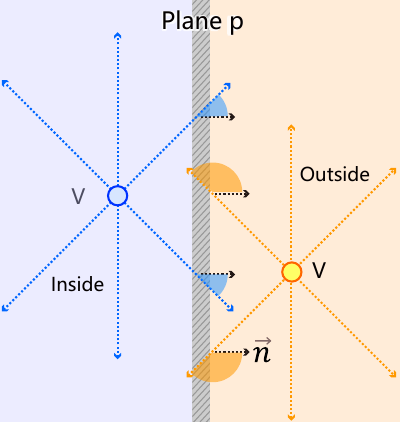} 
	\label{fig:example} 
	\vspace{-0.15cm}
\end{wrapfigure}

\paragraph{\textbf{Initial Segmentation}} 
To initiate structural segmentation, the process begins by detecting primary planes through region growing, using the $\alpha$ shape~\cite{alpha_shape} to define the region for each detected plane. Once these planes are identified, a voxel-based visibility analysis is conducted to group strongly correlated planes into plane-groups. 
	
\begin{figure}[t!]
    \centering
    \includegraphics[width=\linewidth]{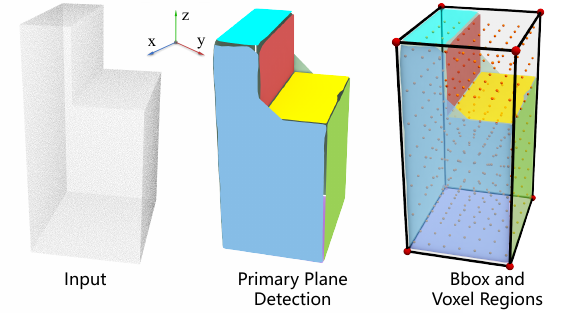}
    \caption{Visualization of primary planes and Bbox with voxel samples.}
    \label{fig:voxelBbox}
\end{figure}

The procedure is outlined as follows:
	1) Bounding Box Sampling: As illustrated in Fig.~\ref{fig:voxelBbox}, we start by computing a bounding box (Bbox) for the detected planes and divide it uniformly into $n_s \times n_s \times n_s$ voxels, each with a sampled centroid.
2) Ray Shooting: From each voxel centroid $v$, $n_r$ rays are emitted, uniformly sampled on the XY plane. These rays are used to determine the planes surrounding the voxel. Specifically, the first plane intersected by a ray is denoted as $p$. If $v$ is inside $p$ (assuming the normal of the plane points outward), the interaction is counted as a valid hit.
3) Visibility Analysis: For each voxel $v$, the visibility information regarding the planes is recorded in an $n_p$-dimensional vector $Hitlist_v$, where $n_p$ is the number of planes. Each entry in $Hitlist_v$ records the number of valid hits from all rays emitted from voxel $v$. 4) Voxel Validation: A voxel with valid hits less than half of the total emitted rays ($n_r$) suggests that the voxel might be outside of the building, leading to its invalidation.
5) Grouping Planes into Plane-Groups: For each valid voxel $v$, all planes with non-zero entries in $Hitlist_v$ form a plane-group. This group represents a subset of planes that define the characteristics of the space occupied $v$, encapsulating essential spatial features for further architectural analysis.

By following these steps, the segmentation yields groups of interrelated planes, termed plane-groups, which are foundational for subsequent levels-of-detail analysis in architectural modeling.

\begin{figure}[t]
    \centering
    \includegraphics[width=\linewidth]{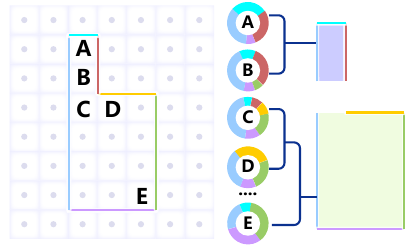}
    \caption{An instance of structural segments aggregation. Plane-groups A, B, C, D, and E are concentrated based on related voxel regions. According to the overlap in $Hitlist$, interrelated planes (blue, cyan, and brown lines 
in the upper right corner) is detected to aggregate structures (A, B), then a structural segment (corresponds to the pale purple area) is generated.}
    \label{fig:segments-generation}
\end{figure}

\paragraph{\textbf{Segments Aggregation}} Once the plane-groups are obtained, we further aggregate them to generate structural segments as depicted in Fig.~\ref{fig:segments-generation}. Each segment represents a collection of planes surrounding a standard component composed of multiple voxels. The aggregation mainly involves organizing the plane-groups to obtain more compact and concise information, corresponding to the architectural LOD. Let $g_a$ and $g_b$ represent the two plane-groups enclosing voxels $V_a$ and $V_b$, respectively. The total valid hits of a plane-group is calculated by
\begin{equation}
hSum(g,V) =  \sum\limits_{v\in V}^{}\sum\limits_{p\in g}^{}Hitlist_v[p],
\end{equation}
where $g$ represents a plane-group, $p$ is a plane belong to $g$, $V$ is the related voxels. 
Then, the pre-condition for merging can be written as $hSum(g_a\cap g_b,V_a) > 0.5 \times hSum(g_a,V_a)$ 
and $hSum(g_a\cap g_b,V_b) > 0.5 \times hSum(g_b,V_b)$. It 
means that the merging of two plane-groups should satisfy a preset overlapping ratio. We iteratively assess the potential merging of plane-groups in pairs and update the merged plane-group as $g_{ab} =g_a \bigcup g_b$,  continuing this process until no further pairs are available for merging. A merging queue is maintained to keep current segments. Only the adjacent segments are checked in the queue which avoid exhaustive searching. Finally, we achieve the merged plane-groups to be structural segments. It should be noticed that a single plane could be part of multiple structural segments. 
To achieve disjointed segmentation results, 
we assign each plane to the structural segment with largest validly hit number. These structural segments serve as primitives for subsequent co-analysis.

\begin{algorithm}[t]
    \caption{Structural Segments Generation}\label{algo1}
    \KwIn{Building model, denoted as $m$}
    \KwOut{A set of segments of $m$, denoted as $S$}    
    $P\leftarrow$ RegionGrowing($m$)\;
    
    \textcolor{blue}{// Handle alcove structures}

    Reverse the orientation of planes $p\in P$ with an area less than $A_{\epsilon}$
    
    $G\leftarrow$ InitialSegmentation($P$)

    $S_a\leftarrow$ SegmentsAggregating($G$)
    
    \textcolor{blue}{// Handle closed structures}

    Recover the orientation for all planes $p\in P$
       
    \For{$p\in S_a$}{
        $P\leftarrow P/\{p\}$
    }
        
    $G\leftarrow$ InitialSegmentation($P$)

    $S_c\leftarrow$ SegmentsAggregating($G$)

    $S\leftarrow S_a \cup S_c$
\end{algorithm}

\subsection{Joint Structural Analysis} \label{co-analysis}
Based on the structural segments, Co-LOD performs co-analysis to ensure semantic consistency in LOD generation across different buildings. To implement the scheme of joint structural analysis, we have designed a two-step method. In the first phase, we identify segments belong to LOD0 through joint analysis of structural segments. It is used to define the initial main structure of the building. Next, we employ spectral clustering to jointly analyze segments beyond LOD0, assigning them to different layers to generate LODs. In this way, Co-LOD achieves progressive LOD generation layer by layer.

\paragraph{\textbf{Initial LOD0 Generation.}} Firstly, we provide the implementation of LOD0 generation for initial main structure of building representation. A reasonable LOD0 definition should satisfy following requirements:  corresponds to the overall shape, as concise as possible, and keeps semantic consistency. Based on the requirements, we formulate the objective function using three terms, which quantitatively evaluate shape fidelity ($f_r$), simplicity ($f_s$), and semantic consistency ($f_{co}$), respectively. That is:
\begin{equation}
    \begin{aligned}
        &E_{0} = \sum\limits_{i=0}^{n}\Big(f_r(M_i)+\beta f_s(M_i)+ \lambda f_{co}(M_i)\Big),\\
        &\quad\text{s.t.}\quad f_r(M_i)>0.8 \quad\forall i = 1,2,...,n,
    \end{aligned}
    \label{opti_func}
\end{equation}
where $n$ represents the number of buildings for co-analysis, 
$I_i$ represents the set of structural segments associated with the $i$-th building, 
$M_i$ represents the set of structural segments identified as the LOD0 of $I_i$,
$\beta$ and $\lambda$ are two parameters to tune the influence of $f_s$ and $f_{co}$. Higher $E_{0}$ score corresponds to a better LOD0 structure that balances the three requirements. It can be modeled as a binary linear programming problem, which determines whether a segment belongs to LOD0.

Intersection over Union (IoU) between $M$ and $I$ is used to define the shape fidelity term $f_r$. It can be computed as 
\begin{equation}
    f_r(M) = IoU(M,I) = IoU(P(M),P(I)),
    \label{frfunc}
\end{equation}
where $P(X)$ represents voxel centroids enclosed by 
the structural segments collection $X$. To satisfy the second requirement of concise representation, the simplicity term $f_s$ is introduced by constraining the area of planes, computed as
\begin{equation}
    f_s(M) = -area(M)/area(I),
    \label{fsfunc}
\end{equation}
where $area(M)$ represents the total area of planes contained in $M$, $area(I)$ represents the total area of planes contained in $I$. By maximizing the term, geometric structures of LOD0 can be greatly simplified.

\begin{figure}[t]
    \centering
    \includegraphics[width=1\linewidth]{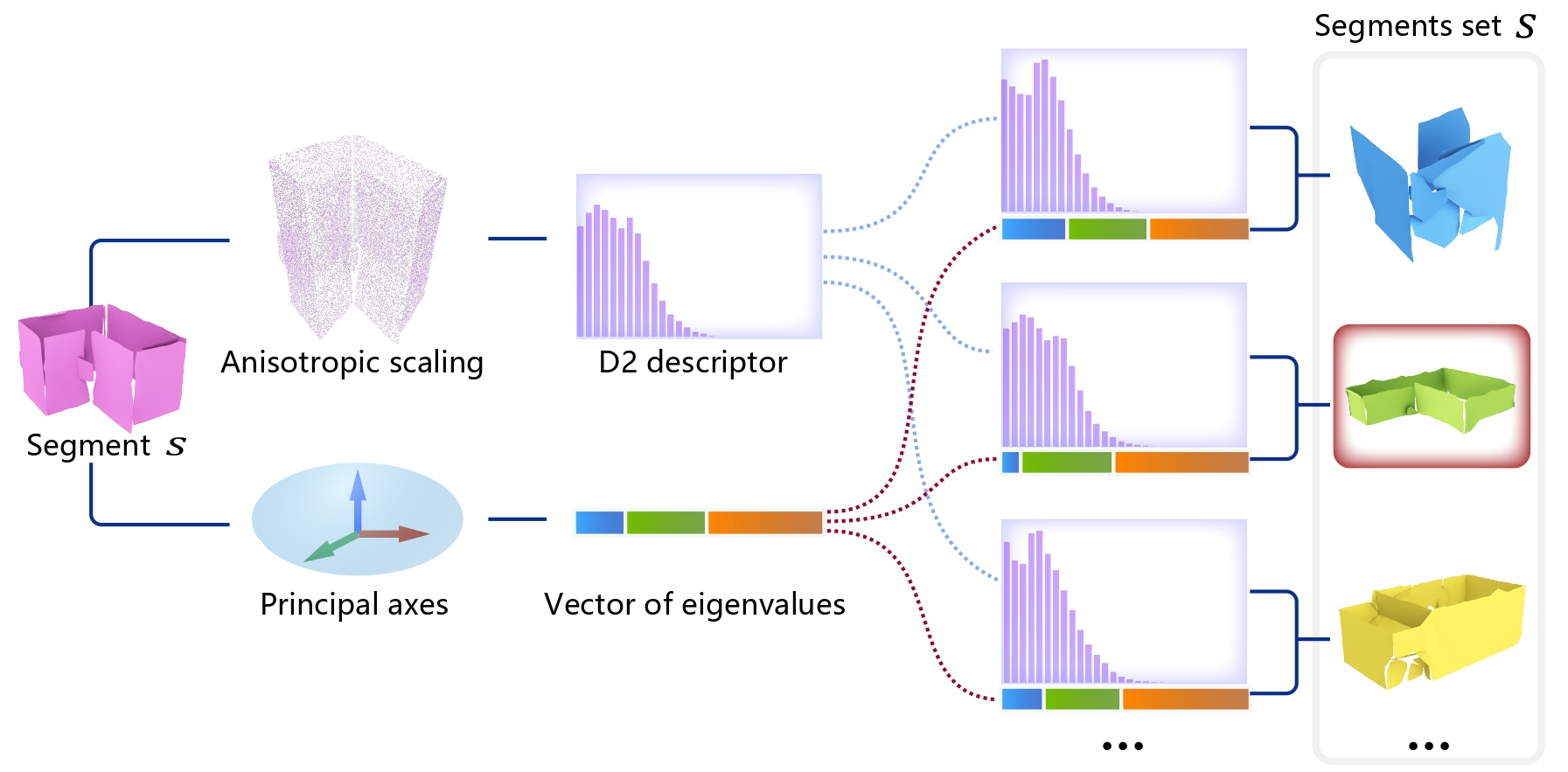}
    \caption{Illustration of segment-based similarity. With the D2 descriptor-based measurement and principal axes-based eigenvalue analysis, different segments can be quantitative compared, which considers shape and scale differences at the same time.}
    \label{fig:similarity}
\end{figure}

\begin{figure}
    \centering
    \includegraphics[width=\linewidth]{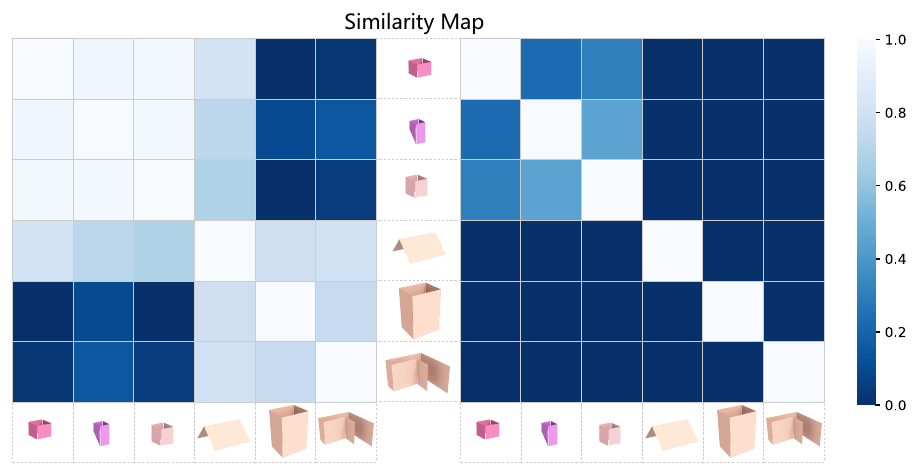}
    \caption{Structural segment-based similarity map by $Dis_{set}$ (left) and $Sim_{co}$ (right). The latter one significantly distinguishes the difference between segments.}
    \label{fig:sim-matrix}
\end{figure}

To address the semantic consistency, we first define the measurement between two segments, represented as
\begin{equation}
    dis(s,s') = \eta \Vert d2(norm(s)) - d2(norm(s')) \Vert_2 + scale_{d}(s,s'),
\end{equation}
\begin{equation}
scale_d = \| v_e(s)-v_e(s')\|/(\| v_e(s)\|+\| v_e(s')\|),
\end{equation}
where $dis$ is a function used to measure the distance between segments $s$ and $s'$. A smaller distance implies a higher similarity between the two segments.
We define the distance with two components, shape distance and scale distance $scale_{d}$. The $norm(s)$ represents the process of making $s$ isotropic through anisotropic scaling~\cite{dis}, 
$d2(x)$ denotes the $D2$ descriptor~\cite{D2} for $x$, and if we denote $\boldsymbol v_{e}(s)$ as the vector of principal 
eigenvalues of the covariance matrix of $s$, 
then $scale_d$ is defined as the normalization of $L1$-distance based on $\boldsymbol v_{e}(s)$ and $\boldsymbol v_{e}(s')$. For intuitively illustrate such similarity, we show an instance in Fig.~\ref{fig:similarity}. According to the segment-based similarity, we perform a cross-analysis of segments on different buildings, formulated as
\begin{equation}
    Dis_{set}(s,S) =\min\limits_{s'\in S}dis(s,s'),    
\end{equation}
\begin{equation}
    Sim_{co}(s,S) =e^{-Dis_{set}(s,S)},
    \label{sim_func}    
\end{equation}
where $Dis_{set}$ represents the distance between $s$ and a segment group $S$, $Sim_{co}$ represents the cross similarity based on $Dis_{set}$, it is improved by logarithmic operation for discrimination (Fig.~\ref{fig:sim-matrix}). Once the cross similarity is defined, the consistency term $f_{co}$ can be computed. It considers the mutual similarities between different segments according to the definition of LOD~\cite{cityGML} and real-life scenario
(segments beyond LOD0 often comprise elements such as windows, fences, rooftop water tanks, air conditioning units, 
and other components that are not individually designed). 
To formulate $f_{co}$, we encourage segments that match well with others to be removed from LOD0 since they represent repetitive structures. In practice, $f_{co}$ is defined as
\begin{equation}
    \begin{aligned}
        f_{co}(M)=  \sum_{s}\sum_{M',I'}Sim_{co}(s,I'-M'),\\
        s\in I-M, M'\in \mathcal{M}-M, I'\in \mathcal{I}-I,
    \end{aligned}    
\end{equation}
where \begin{math}\mathcal{M}\end{math} and $M$ are collections of LOD0-based segments corresponding to all architectural models and current model, \begin{math}\mathcal{I}\end{math} and $I$ are collections of segments from all architectural models and current model.
$I-M$ represents segments beyond LOD0 of current model. By using set operations, we compare segments in certain range by cross similarity $Sim_{co}$ for co-analysis of LOD0. 

\paragraph{\textbf{Spectral Clustering.}} 
Through the optimization of $E_0$ in Eq.~\eqref{opti_func}, 
we obtained an accurate LOD0 architectural model.
To further assign more detailed segments into their appropriate LODs, we employed spectral clustering to simultaneously analyze these segments. Based on the cross similarity $Sim_{co}$, we constructed the similarity matrix $M_s$ to measure the similarity between each pair of segments. Subsequently, we compute the normalized Laplacian matrix $L_{norm}$, represented as:
\begin{equation}\label{Laplacian}
    L_{norm}=M_D^{-1/2}(M_D-M_s)M_D^{-1/2}, M_{D_{ii}}= \sum\limits_j^{}M_{s_{ij}},    
\end{equation}
where $M_s$ is the similarity matrix constructed by cross similarity $Sim_{co}$, $M_{s_{ij}} = Sim_{co}(s_i,s_j)$, $M_D$ is the related diagonal matrix. Inspired by spectral clustering~\cite{Spectral_Clustering}, we use the eigenvectors corresponding to the first $l_n$-1 smallest eigenvalues of $L_{norm}$ as features for segments and cluster them by $k$-means method. Then, such segments are clustered into $l_n$-1 classes, corresponding to $l_n$-1 different LOD levels, $l_n$ corresponds to the required LOD layers. In practice, for datasets featuring finely scanned models, we typically set $l_n=3$ as shown in Fig.~\ref{fig:D4models} and Fig.~\ref{fig:CH_results}. Conversely, for collections with roughly scanned models, we set $l_n=2$, as shown in Fig.~\ref{fig:teaser} and Fig.~\ref{fig:yrs_results}. Clusters with smaller average volume are assigned to higher LOD levels, indicating that they represent finer structures. A higher layer of LOD is formed by the planes corresponding to segments smaller than or equal to the current 
LOD layer with adjacent planes that don't belong to any segments. In this way, we utilize such planes to construct more refined and comprehensive polygonal mesh.

\begin{figure}[t]
    \centering
    \includegraphics[width=0.9\linewidth]{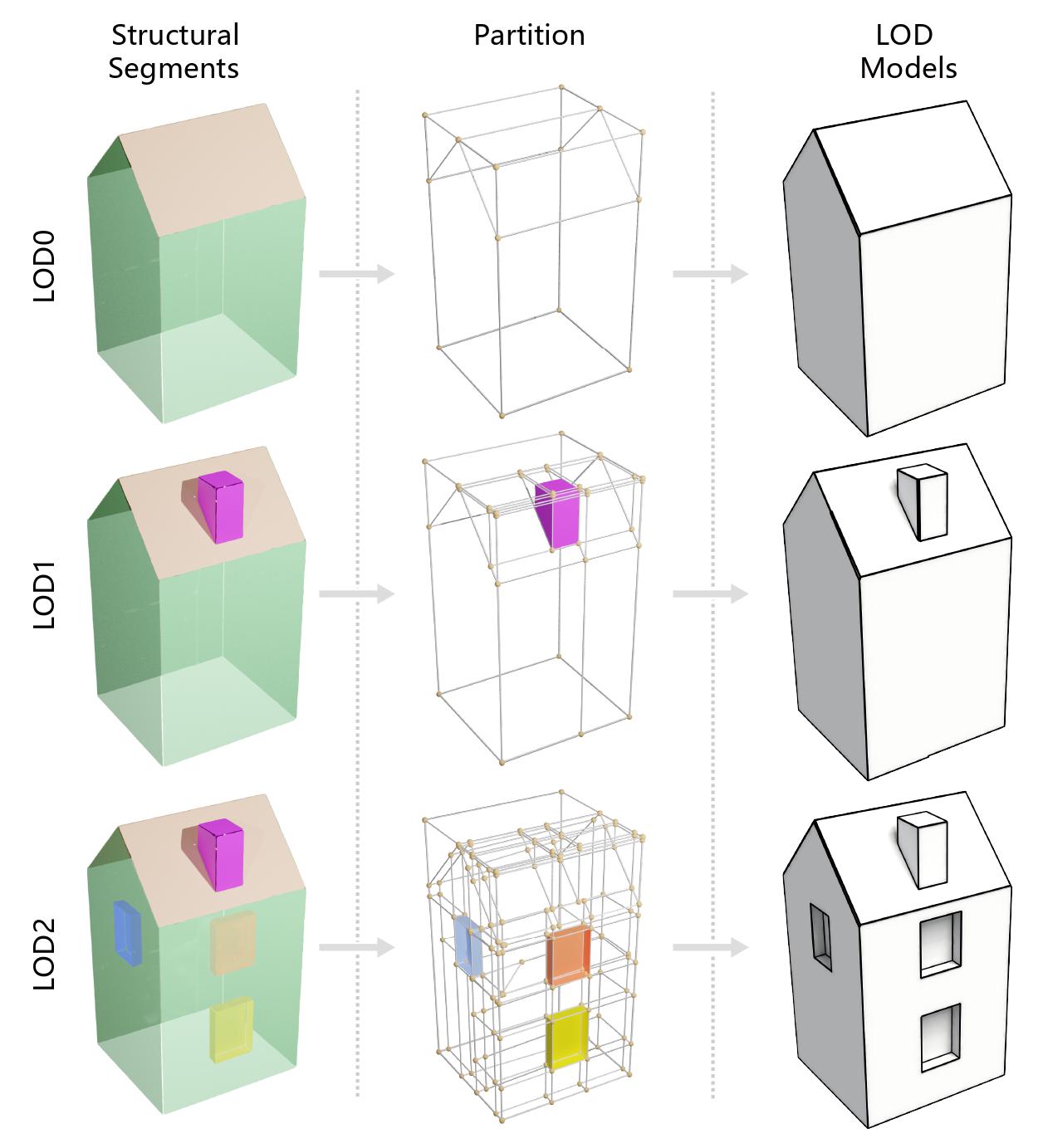}
    \caption{Illustration of polygonal mesh extraction with different LOD layers. By partitioning scheme with layered structural segments, accurate LOD generation can be implemented.}
    \label{fig:sigMesh}
\end{figure}

\subsection{Polygonal Mesh Extraction} \label{Mesh-Extraction}
To achieve the desired polygonal meshes, we utilize a space partitioning approach followed by convex hull inside/outside calibration to extract polygonal meshes based on layered segments. Specifically, planes belonging to specific LOD layers undergo Binary Space Partitioning~\cite{BSP} to divide the architectural model's bounding box into convex hulls. Subsequently, we determine the interior or exterior attributes of each convex hull based on whether more than half of the rays see the convex hull as interior. That is:
\begin{equation}
In(C)=\left\{\begin{array}{l}1, h_v(C) > 0.5n_r\\0, other\end{array}\right.,
\label{label_io}
\end{equation}
where $In(C)$ is the index of interior ($In(C)=1$) or exterior ($In(C)=0$) attribute, $C$ is the centroid of the related convex hull that also can be regarded as a sub-space, $h_v(C)$ is the total valid hit count from $n_r$ rays uniformly emitted from $C$. Based on all interior convex hulls, the largest internally connected component forms the watertight space, which produces the polygonal mesh for current layer of LOD. An instance is shown in Fig.~\ref{fig:sigMesh}. Unlike traditional methods that directly use detected planes for polygonal mesh reconstruction, Co-LOD utilizes segment co-analysis to guide clustering and concentrating related planes, which ensure precise and controllable LOD generation with semantic consistency. In experiments, we illustrate the performance of Co-LOD in practical urban scenes.

%% file: src/04-results.tex
\section{Experimental Results}
\label{sec:results}

\begin{figure*}
	\centering
	\includegraphics[width=\linewidth]{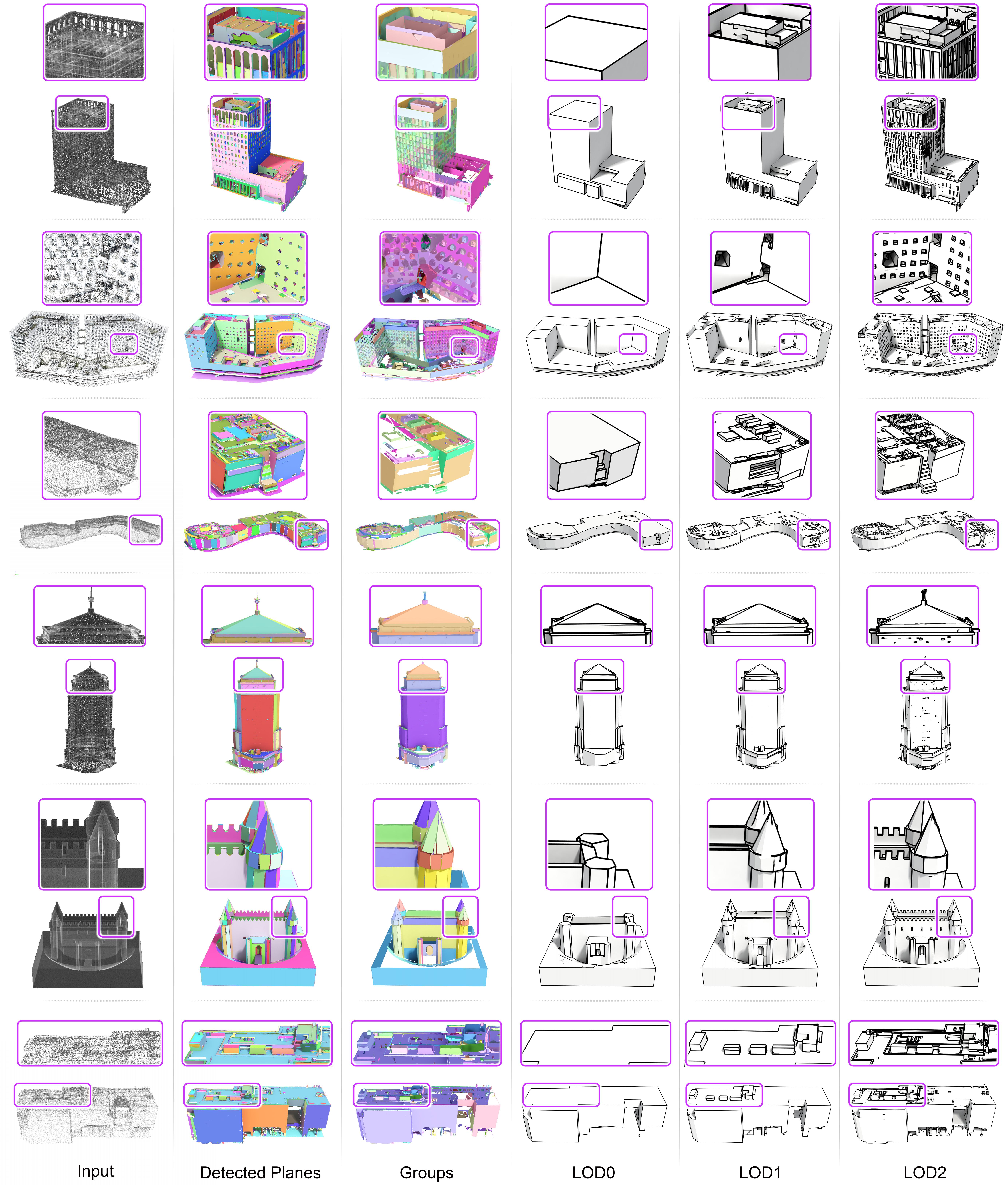}
	\caption{LOD generation results for \textit{Composite Scene}, demonstrating Co-LOD's ability to handle both diverse styles and delicate structures.}
	\label{fig:D4models}
\end{figure*}

\begin{figure*}
	\centering
	\includegraphics[width=0.97\linewidth]{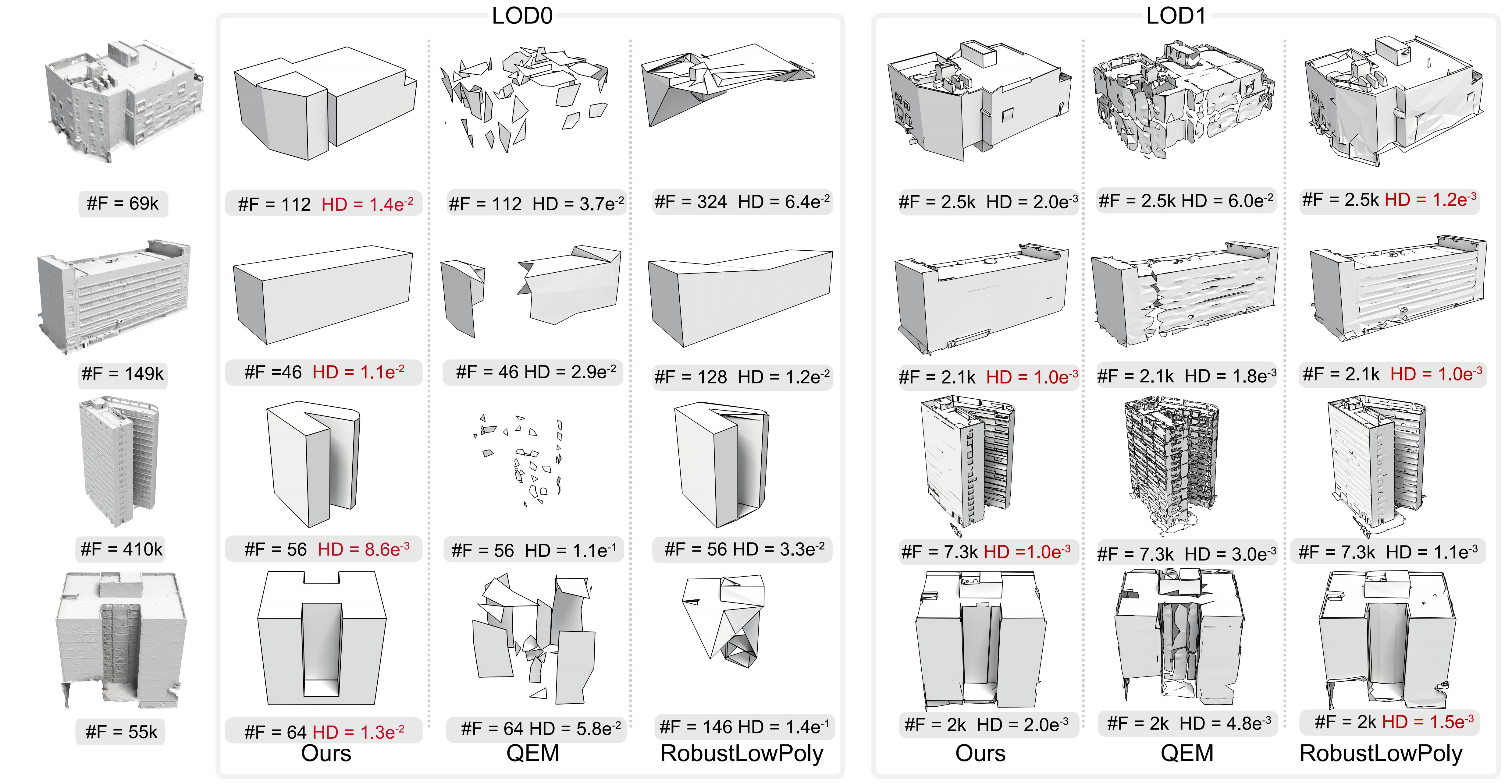}
	\caption{Comparisons with QEM and RobustLowPoly for LOD generation. Co-LOD achieves better balance between simplicity (\#F: face number) and accuracy (HD: Hausdroff distance between the raw input and related reconstruction result).}
	\label{fig:compare_qem_RobustLowPoly}
\end{figure*}

We evaluate the performance of Co-LOD in the context of urban scene-based architectural LOD generation tasks.  All experiments were conducted on a machine equipped with an Intel i9-13900K processor, 128GB RAM, and an RTX4090 graphics card, running Windows 10 and Visual Studio 2022. For linear algebra operations, basic geometry processing, and numerical optimization, we utilized public libraries such as Eigen~\cite{eigenweb}, CGAL~\cite{CGAL}, and LPSolve. Comparative analysis between Co-LOD and several classical solutions, 
as well as single building processing and ablation studies, are used to comprehensively assess the effectiveness of Co-LOD.

\paragraph{\textbf{Datasets}} 
To  thoroughly evaluate the effectiveness of our method, we investigated building datasets published post-2020, including Toronto-3D~\cite{tan2020toronto3d}, SensatUrban~\cite{hu2022sensaturban}, SUM~\cite{sum2021}, STPLS3D~\cite{chen2022stpls3d},  
InstanceBuilding~\cite{instance3d}, and UrbanBIS~\cite{UrbanBIS}. We conducted experiments on the datasets with normal information, totaling 421 architectural data samples of various styles and captured from diverse locations. These models were reconstructed using UAV photogrammetry, which introduced inherent noise and incomplete information. We organized the samples into seven scenes. Following the approach in LowPoly~\cite{low-poly}, we compiled the statistics for the scenes used in Table~\ref{table:input}. As Co-LOD takes a collection of building as input, the seven building scenes are individually fed into the pipeline for co-analysis.

\input{tables/input.tex}

\begin{figure}
	\centering
	\includegraphics[width=\linewidth]{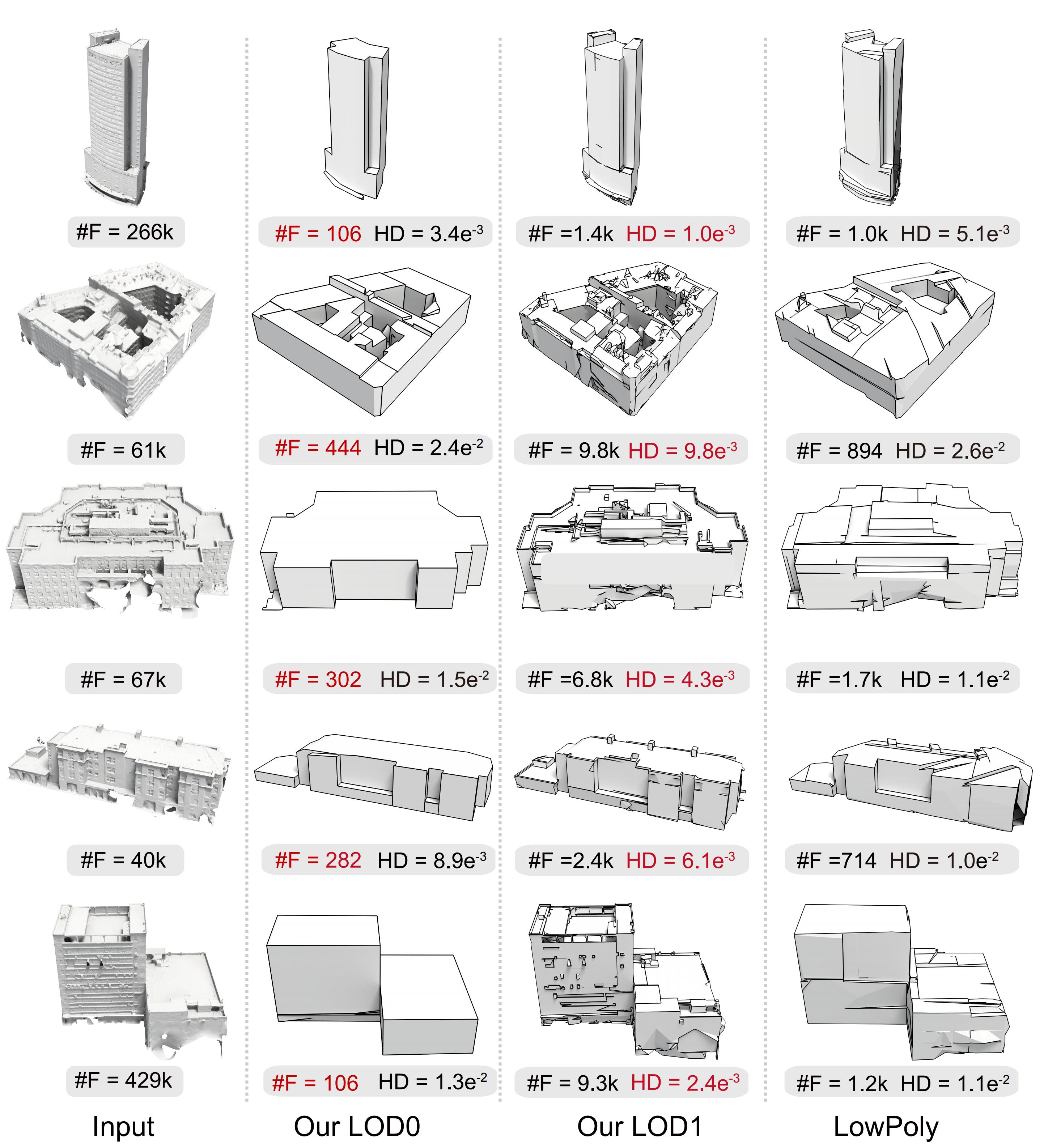}
	\caption{Comparisons with LowPoly for LOD generation. LowPoly fails to keep some main structures. Co-LOD achieves more concise and clear main structures.}
	\label{fig:compare_lowpoly}
\end{figure}

\begin{figure*}
	\centering
	\includegraphics[width=\linewidth]{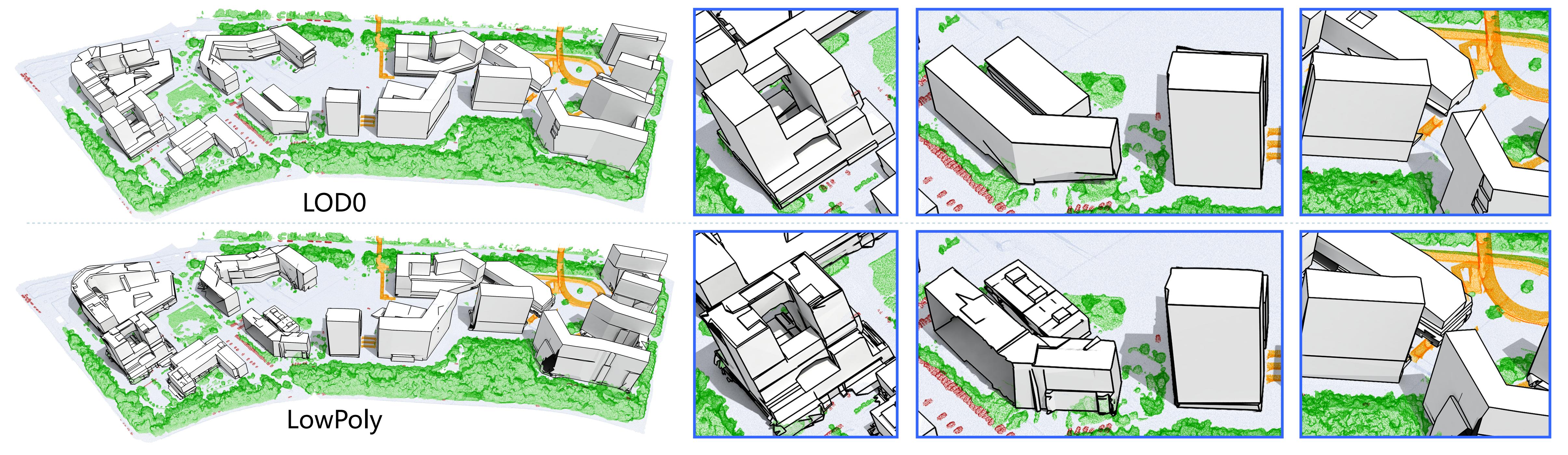}
	\caption{Comparisons with LowPoly for LOD generation using \textit{Research Center}. The zoomed-in views highlight the notable inconsistencies in the amount of details used by LowPoly to model different buildings.}
	\label{fig:CH_compare_Lowpoly}
\end{figure*}

\input{tables/Comparison_statistics.tex}

\paragraph{\textbf{Metrics \& Configurations}} 
To evaluate the performance of the LOD generation methods, we employ Hausdorff distance (HD) and Light Field distance (LFD)~\cite{lfd} to measure geometric and visual errors. We also report success rate ($r_s$) and user study at both individual and scene levels to reveal the practicality.
For parameter configurations of Co-LOD, we set $n_s=150$, $n_r=100$, $A_\epsilon=80m^2$, $\beta = 0.3$, $\lambda = 0.4$, $\eta = 4.0$ as the default specification. The rationality of parameter selection are discussed in the subsequent sections. For the plane detection, we employ the Region Growing algorithm~\cite{RegionGrowing} provided by CGAL. The distance and angle-based thresholds are set to 0.2m and 30°, respectively, and the minimum region size requiring 20 inside points. The results generated using Co-LOD for some instances are shown in Fig.~\ref{fig:D4models}.

\begin{figure*}
	\centering
	\includegraphics[width=\linewidth]{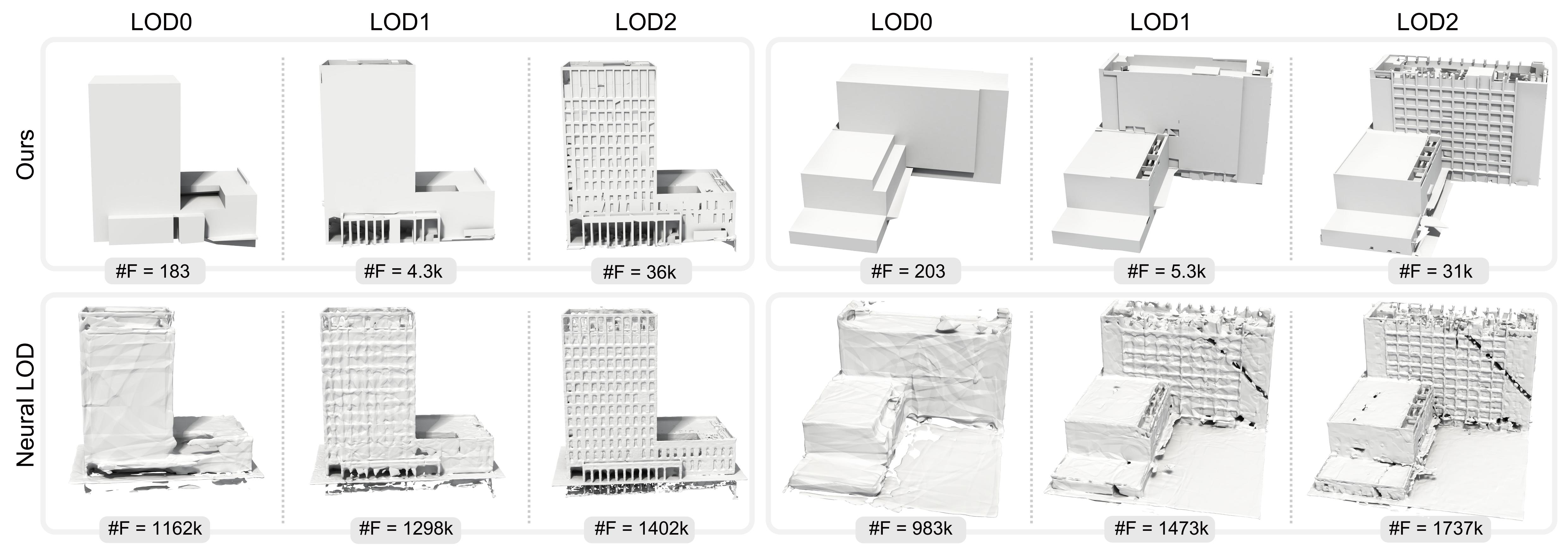}
	\caption{Comparisons with NeuralLOD. The face number of NeuralLOD-based results are huge even in models with LOD0 layer. Co-LOD effectively reduces the face number while maintaining accurate geometric structures in different LOD layers.}
	\label{fig:NeuralLOD}
\end{figure*}

\subsection{Comparisons}
To validate the effectiveness of Co-LOD, we conducted comparisons with several mainstream approaches, including QEM~\cite{QEM}, RobustLowPoly~\cite{RobustLowPoly}, LowPoly~\cite{low-poly}, NeuralLOD~\cite{neuralLOD}, PolyFit~\cite{nan2017polyfit}, and KSR~\cite{KSR}. These comparisons were based on the previously mentioned metrics and included user studies at both individual and scene levels. Additionally, we presented the generated LOD models visually to demonstrate the hierarchical consistency advantage of Co-LOD.

\paragraph{\textbf{Comparisons with QEM and RobustLowPoly.}} 

We conducted comparative experiments using QEM, RobustLowPoly, and Co-LOD, focusing on geometric and visual error optimization. Due to limitations in the ability of the comparison methods to implement accurate LOD control, we simplified the input facets to simulate the function of Co-LOD generation for a fair assessment. As demonstrated in Fig.~\ref{fig:compare_qem_RobustLowPoly}, QEM demonstrates limitations in preserving architectural integrity, particularly struggling to retain sharp features during LOD generation, which results in poor visual effects. RobustLowPoly, while preserving basic geometric details more effectively, still falls short in ensuring the consistency of building structures.  Additionally, it lacks the capability to provide structural completion and to smooth noisy planes. Moreover, both methods struggle to maintain the main structure when simplified to a very low number of faces, leading to lower geometric and visual fidelity and inconsistencies within the LOD0 layer.

Benefited from the co-analysis, Co-LOD excels in maintaining structural integrity and achieving better semantic consistency with fewer planes. These advantages are evidenced by quantitative evaluations of HD and LFD, as detailed in Table~\ref{table:Comparison}. 

\begin{figure}[t]
	\centering
	\includegraphics[width=\linewidth]{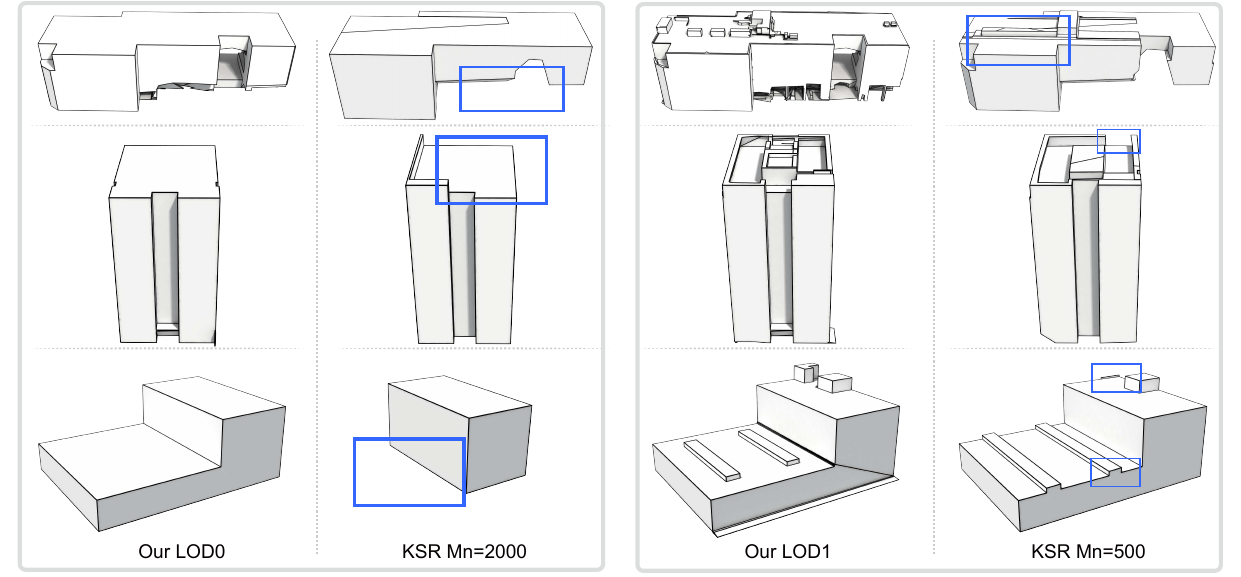}
	\caption{Comparisons with Alternative Pipeline. With the parameter changing, KSR can generate different levels of geometric structures. However, the main structure can not be guaranteed which poses significant drawback for geometric and semantic consistency.}
	\label{fig:plane_detection_params}
\end{figure}

\begin{figure*}[t]
	\centering
	\includegraphics[width=\linewidth]{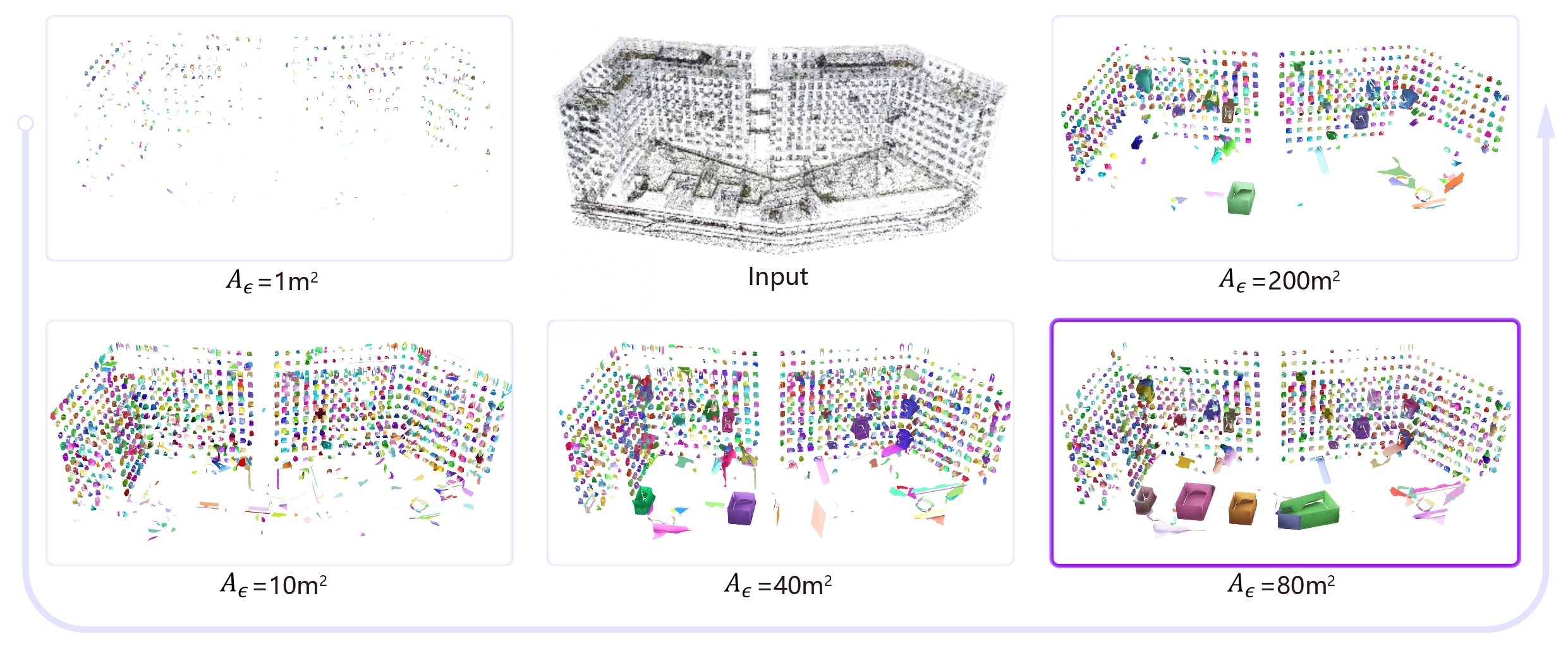}
	\caption{The impact of \(A_{\epsilon}\) on alcove structure detection. While a larger \(A_{\epsilon}\) facilitates the detection of larger alcove structures, excessively increasing \(A_{\epsilon}\) to \(200m^2\) can impede detection by causing the inversion of large planes.}
	\label{fig:A_eps}
\end{figure*}

\begin{figure}[t]
	\centering
	\includegraphics[width=\linewidth]{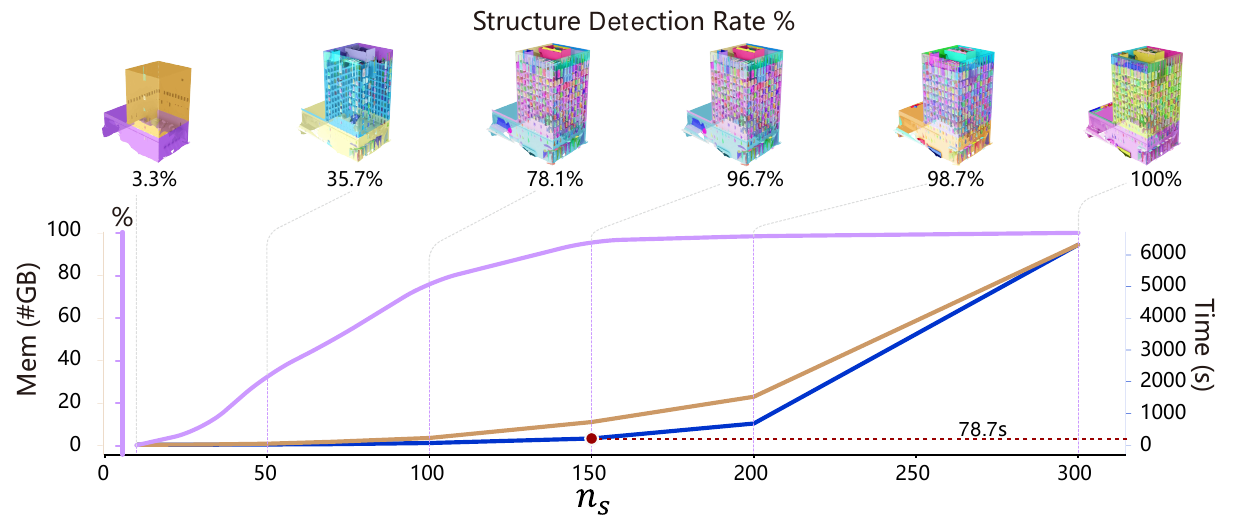}
	\caption{Relationship between $n_s$ and computational cost, benchmarked at $n_s = 300$ for SDR calculations. Increasing $n_s$ improves detection of fine structures but incurs a non-linear rise in computational costs.}
	\label{fig:ns_ablation}
\end{figure}

\begin{figure}[t]
	\centering
	\includegraphics[width=\linewidth]{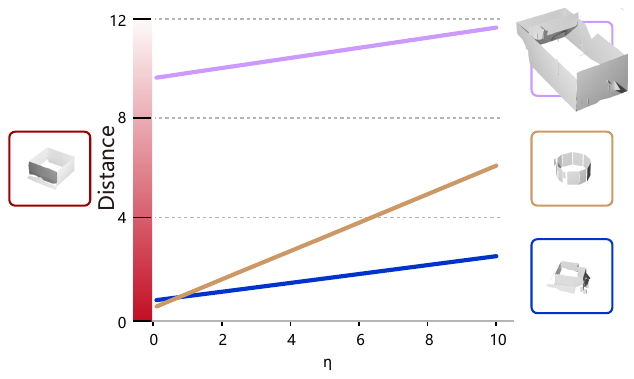}
	\caption{Impact of the shape-scale balance weight $\eta$ on segment similarity: Increasing $\eta$ shifts the metric emphasis from scale to shape similarity.}
	\label{fig:lambda_d}
\end{figure}

\paragraph{\textbf{Comparisons with LowPoly.}}
LowPoly can construct the main structure of an architectural model, making it a suitable benchmark for evaluating the LOD generation capabilities of Co-LOD. However, it has some limitations in maintaining semantic consistency within the LOD layer, as shown in Fig.~\ref{fig:compare_lowpoly} and Fig.~\ref{fig:CH_compare_Lowpoly}. Different models exhibit varying levels of simplification, and repetitive structures are not consistently removed. In contrast, Co-LOD can generate a more concise and uniform LOD0 layer across various architectural models. For LOD1 generation shown in Fig.~\ref{fig:compare_lowpoly}, Co-LOD produces more accurate geometric structures and better semantic consistency. Further comprehensive results are reported in Table~\ref{table:Comparison}.

\begin{figure*}[t]
	\centering
	\includegraphics[width=0.9\linewidth]{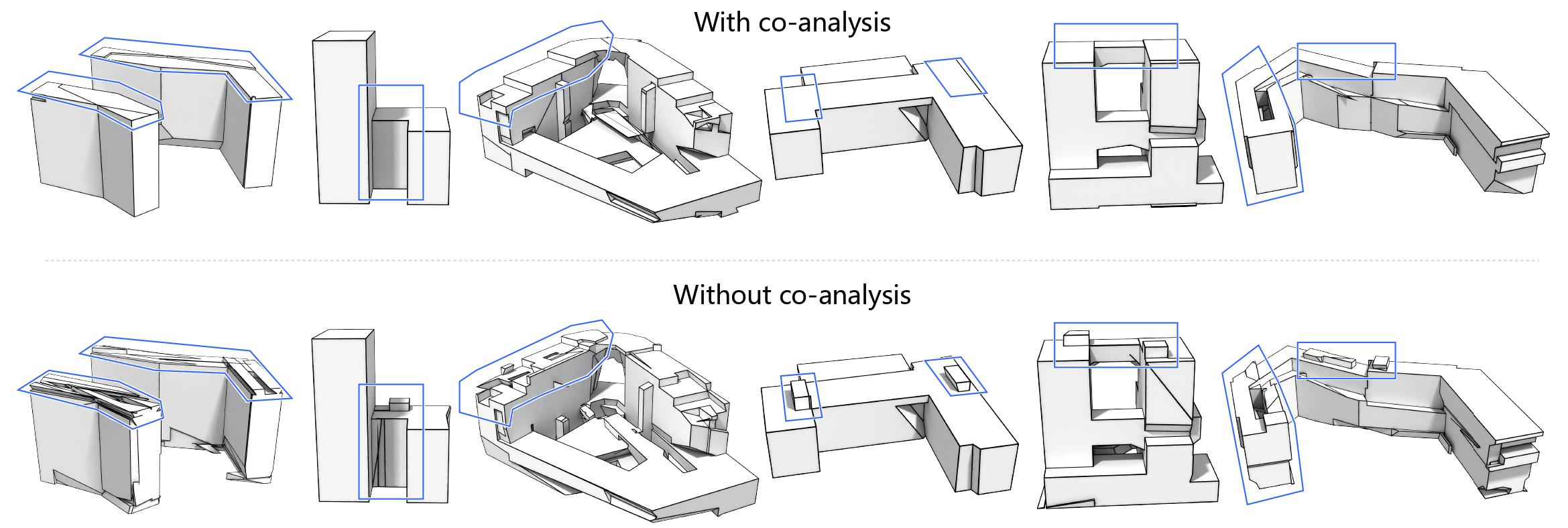}
	\caption{Ablation study on the impact of consistency term $f_{co}$ on LOD0 generation. The top row shows results with the term, where LOD0 exhibits concise and consistent geometric structures. In contrast, the bottom row, without $f_{co}$, displays irregular geometric structures that introduce disturbances.}
	\label{fig:Ablation1}
\end{figure*}

\begin{figure*}
	\centering
	\includegraphics[width=0.95\linewidth]{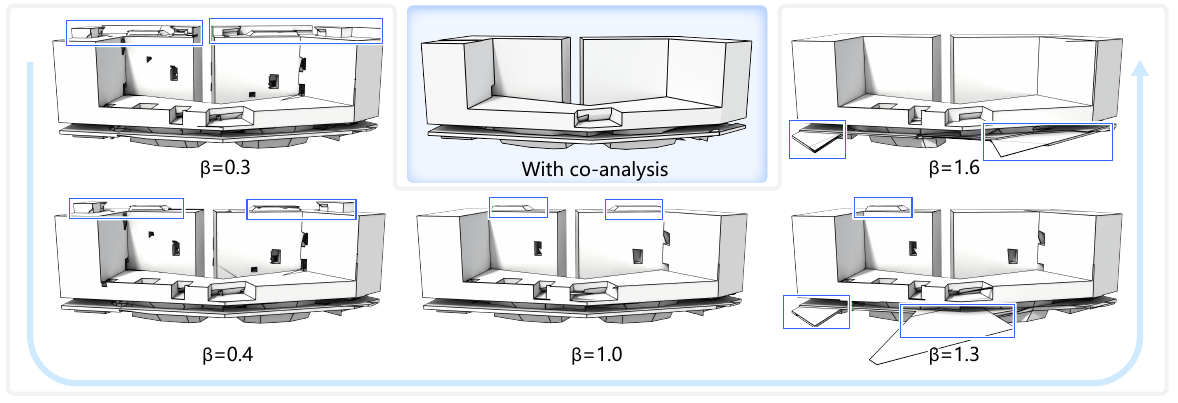}
	\caption{Evaluation of shape term $f_r$ and simplicity term $f_s$ in Co-LOD without function of $f_{co}$. Tuning the ratio between the two terms cannot effectively generate a desired LOD0 model, underscoring the indispensable role of co-analysis in Co-LOD controlling.}
	\label{fig:Ablation2}
\end{figure*}

\paragraph{\textbf{Comparisons with NeuralLOD}}

The advent of deep implicit surface reconstruction has brought new possibilities to the generation of various LOD layers for architectural models, as exemplified by NeuralLOD. In Fig.~\ref{fig:NeuralLOD}, we present a comparison of architectural models generated by both NeuralLOD and Co-LOD, across different LOD layers.
NeuralLOD, particularly at lower LOD layers, tends to smooth out geometric details. While this is consistent with the principles of LOD generation, the numbers of faces needed to represent these smooth surfaces are significantly larger than the ones needed by Co-LOD. In addition, NeuralLOD does not operate as a structured reconstruction scheme and lacks the capability to extract primary planes. Consequently, even at higher LOD layers, Co-LOD demonstrates a significant advantage in terms of simplification, while simultaneously maintaining the accuracy of the geometric structures. This comparison underscores the efficiency and precision of Co-LOD in generating LOD layers for architectural models.

\paragraph{\textbf{Comparisons with PolyFit and KSR}} 

For architectural model reconstruction, classical methods~\cite{Chauve,nan2017polyfit,KSR} focus on detecting primary planes to generate polygonal models. While these methods are not tailored for precise LOD control, they offer some level of accuracy adjustment in reconstructed buildings through parameter tuning. The key parameter is the minimum number of inliers ($Mn$), which influences the scale of detected planes by dictating the minimum points needed for a valid plane. We use KSR~\cite{KSR} as a baseline method and adjust the $Mn$ to simulate LOD control. The reconstructed architectural models shown in Fig.~\ref{fig:plane_detection_params} illustrate that different $Mn$ values for KSR can indeed control geometric details. However, it cannot ensure the integrity of the main structure. Precise geometric details obtained with lower $Mn$ are inconsistent. Conversely, Co-LOD effectively addresses the issue, achieving accurate and uniform LODs. Furthermore, Table~\ref{table:Comparison} quantitatively demonstrates that Co-LOD achieves higher modeling accuracy at similar complexities and handles complex inputs more effectively.

\paragraph{\textbf{User Study}} 

We conducted a user study utilizing all scenes and randomly selected 30 individual models, to evaluate the quality of LOD models generated by various methods at both the individual model and scene levels. This study involved 108 participants who chose the best set of LODs from four groups, each provided by a different method, based on their balance between fidelity and simplicity. Detailed results are presented in Table~\ref{table:Comparison}. Our method was preferred for both individual buildings and scenes, receiving approval ratings of 77.3\% and 81.3\%, respectively. This confirms the effectiveness of our approach in generating LOD models that are both geometrically and visually accurate, as well as aesthetically appealing. For visualizations of the comparison results, please refer to our supplementary materials.

\paragraph{\textbf{Co-LOD for Single Building}}
By default, Co-LOD is used to control LOD generation in scenes with multiple buildings while maintaining semantic consistency. When processing single buildings, Co-LOD can benefit from incorporating an additional database to build prior knowledge. To demonstrate this, we utilized \textit{Composite Scene} to compile a database consisting of 90 segments with various shapes. The experiment, conducted using \textit{Campus}, assessed the effectiveness of co-analysis when supplemented with additional database knowledge. Table~\ref{table:signle_analysis} demonstrates that using a database significantly reduces calculation time, primarily by speeding up the co-analysis stage. However, a downside is that non-primary structures not matched within the constructed database may be erroneously assigned to lower LOD layers, resulting in insufficient simplification. Further details of this implementation can be found in the supplementary materials.

\input{tables/single_anlysis.tex}

\input{tables/nr.tex}

\begin{figure*}
	\centering
	\includegraphics[width=\linewidth]{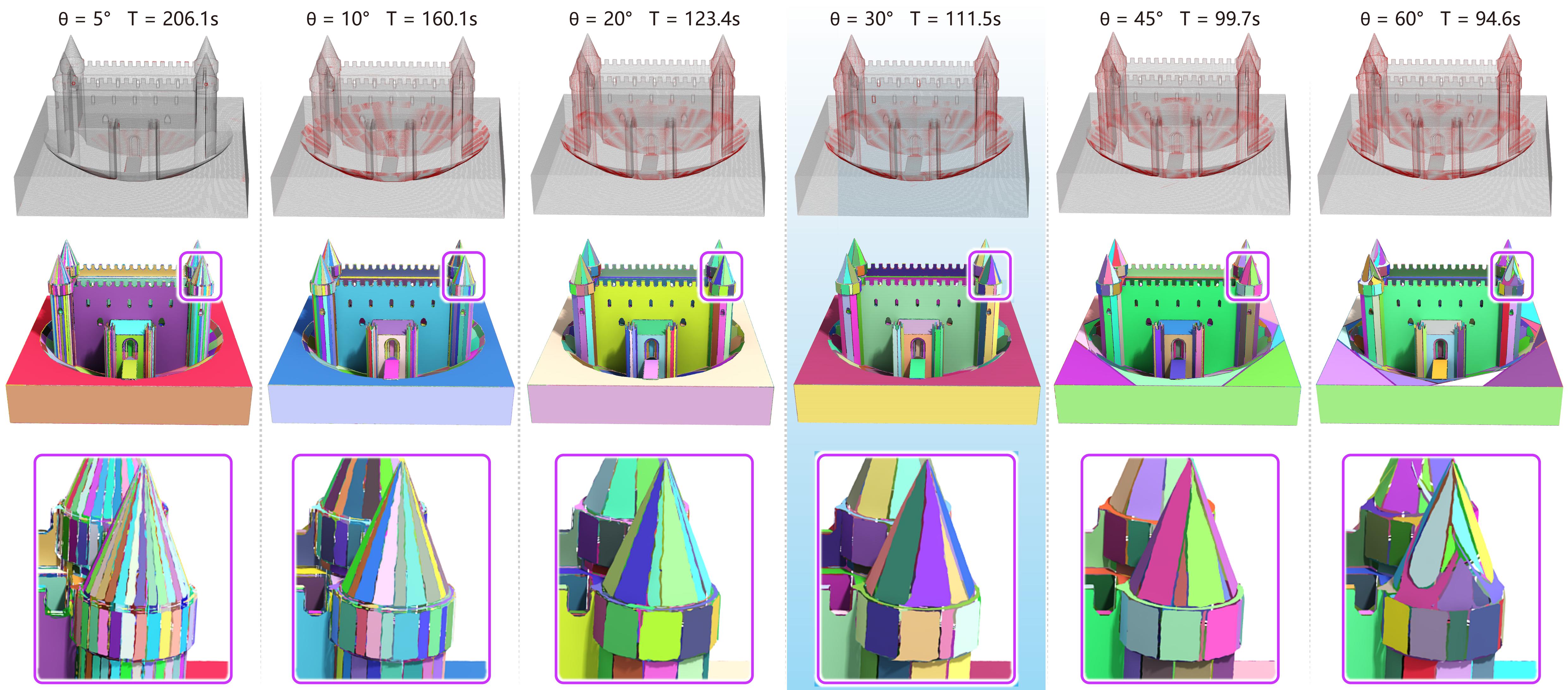}
	\caption{Errors from plane-based surface approximation. Reducing the angle-based threshold ($\theta$) mitigates these errors but increases computational time (T). Deeper reds in the point cloud signify higher reconstruction errors.}
	\label{curve_limitataion}
\end{figure*}

\begin{figure}
	\centering
	\includegraphics[width=\linewidth]{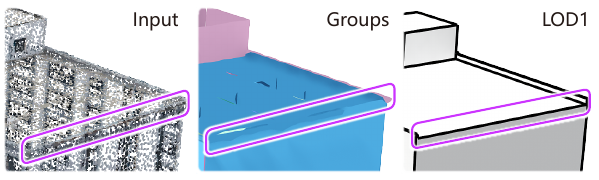}
	\caption{Limitation in detecting finer structures. The inability to extract the circled structure led to inconsistency in the LOD1 model.}
	\label{fine_structure}
\end{figure}

\begin{figure}
	\centering
	\includegraphics[width=0.8\linewidth]{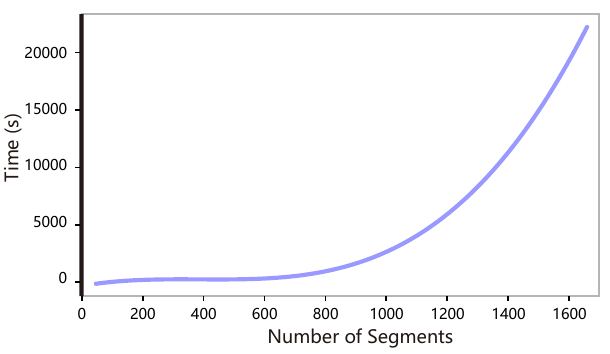}
	\caption{The relationship between the calculation time of the co-analysis stage and the number of segments analyzed simultaneously.}
	\label{fig:Scalability}
\end{figure}

\begin{figure*}
	\centering
	\includegraphics[width=\linewidth]{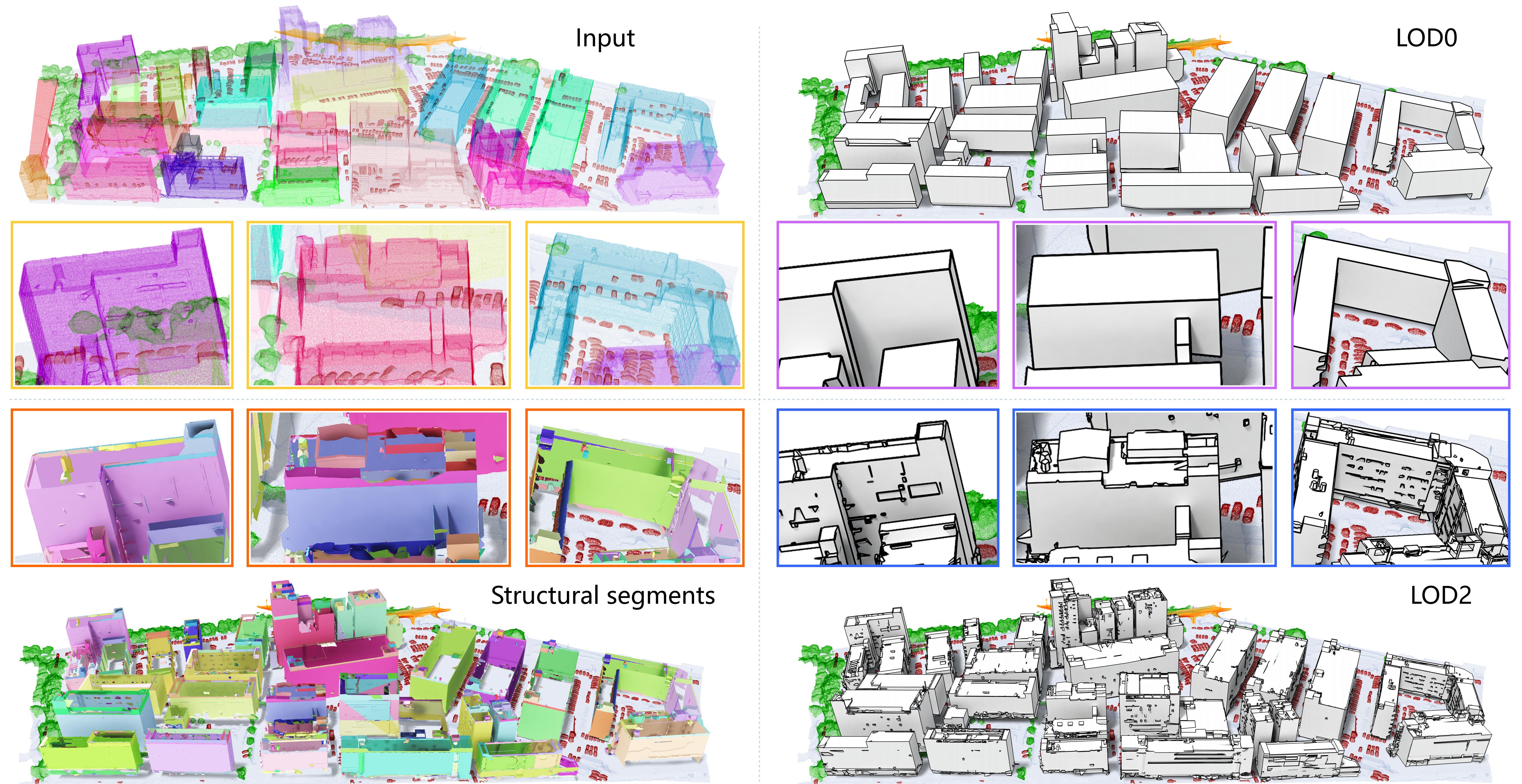}
	\caption{LOD generation results by Co-LOD for \textit{Town}.}
	\label{fig:yrs_results}
\end{figure*}

\begin{figure*}
	\centering
	\includegraphics[width=\linewidth]{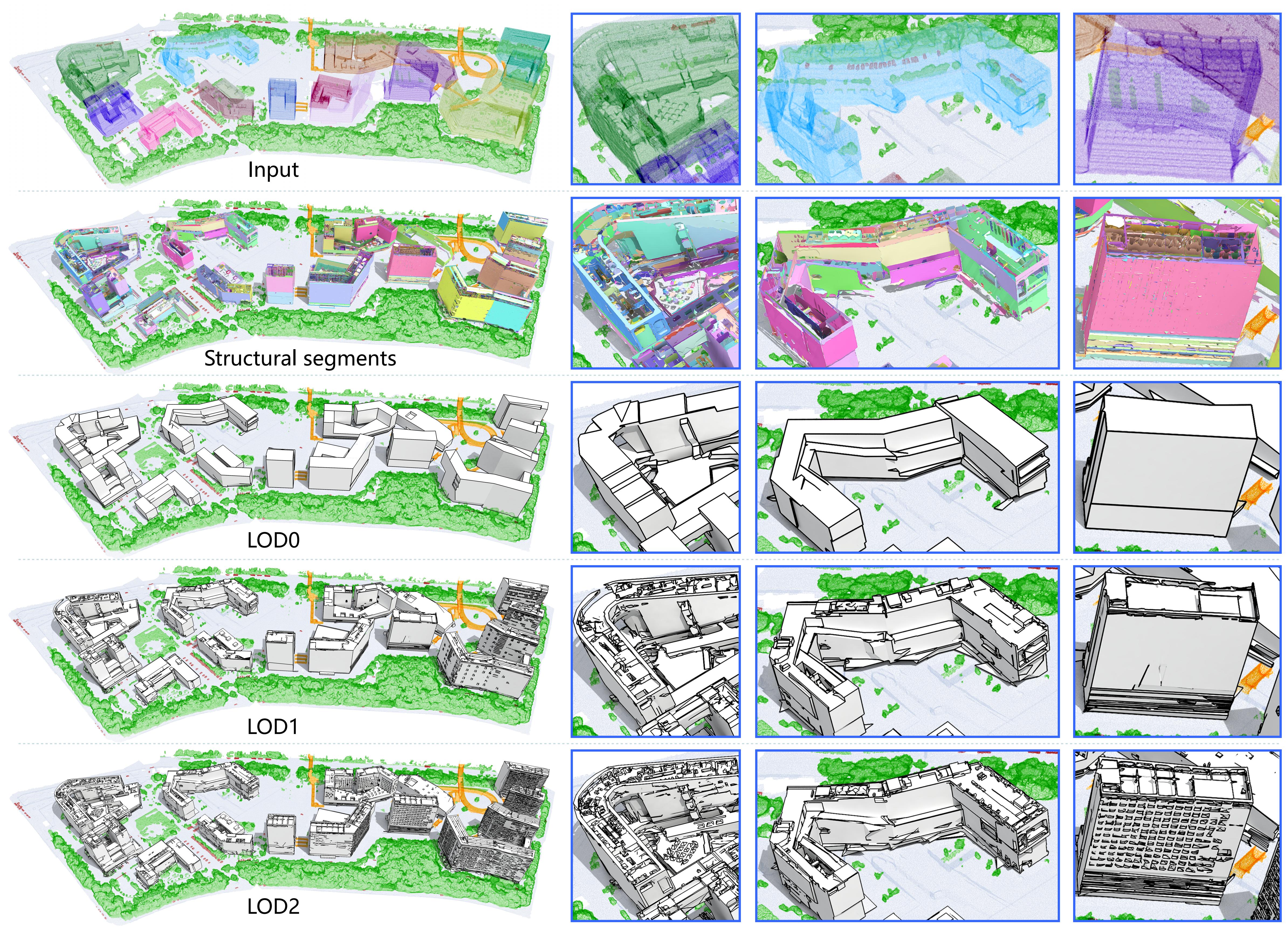}
	\caption{LOD generation results by Co-LOD for \textit{Research Center}.}
	\label{fig:CH_results}
\end{figure*}

\paragraph{\textbf{Efficiency}} 

Across the tested dataset, Co-LOD takes an average of 151.6s to process an input, including stages for structural segment generation (78.7s), co-analysis (51.8s), and polygonal mesh extraction (21.1s). In contrast, comparative methods include KSR, PolyFit, LowPoly, QEM, and RobustLowPoly take 87.3s, 5.1s, 513.3s, 2.1s and 603.2s, respectively. Overall, Co-LOD achieves a better balance in terms of computational efficiency, LOD control, semantic consistency, and practicality. 

\subsection{Ablation \& Limitations}
Co-LOD delivers accurate LOD generation with consistency for architectural models. Given the multiple control parameters and optimization elements within Co-LOD, we conducted an ablation study to elucidate their impacts and discuss some limitations.

\paragraph{\textbf{Ablation Study}} Co-LOD implementation involves various parameters and optimization items that require careful consideration, including alcove structure area $A_\epsilon$, BBox resolution $n_s$, number of Rays Emitted $n_r$, shape-scale balance weight $\eta$, and terms in Eq.~\eqref{opti_func}. 

The selection of \(A_{\epsilon}\) primarily influences the detection of alcove structures, setting the upper size limit of detectable alcoves, as illustrated in Fig.~\ref{fig:A_eps}. Through extensive experimentation, we have set \(A_{\epsilon}\) to 80m\(^2\), which effectively handles the alcove structures contained in our datasets.

The BBox resolution influences structural segmentation, which in turn impacts LOD generation. In Fig.~\ref{fig:ns_ablation}, we explored various $n_s$ values across our dataset and introduced the Structure Detection Rate (SDR) as our custom metric. SDR measures how accurately the Structural Segment Generation algorithm identifies relevant structures. It is calculated as the percentage of segments generated with the current parameter settings compared to those generated using the optimal parameters. Our findings indicate that increasing $n_s$ detects finer structures but also raises computational consumption non-linearly. Therefore, we have set $n_s=150$ to achieve the optimal balance between efficiency and performance. 

The number of emitted rays $n_r$ affects structural segment generation and polygonal mesh extraction, as shown in Table~\ref{table:nr}. This parameter is crucial for analyzing how a voxel is surrounded by planes; fewer rays increase the likelihood of missing surrounding planes, leading to inferior structural segmentation and mesh extraction. Considering that increasing $n_r$ does not lead to non-linear growth in computational costs, we have chosen a larger default value (100) to enhance the robustness and HD-based evaluation. 

Different values of $\eta$ yield varied similarity measurements that influence LOD hierarchy as depicted in Fig.~\ref{fig:lambda_d}. Larger value of $\eta$ prioritizes the shape and smaller one emphasizes the scale. Based on the extensive experiments, we set $\eta = 4.0$ to align shape and scale measurement space.

There are several optimization terms in Eq.~\eqref{opti_func} and their functions should be demonstrated, which take important influence for LOD0 generation. First, we evaluate the function of consistency term $f_{co}$ that provides semantic consistency. In Fig.~\ref{fig:Ablation1}, we compare the results with and without $f_{co}$ in Co-LOD. The $f_{co}$ suppresses the intricate geometric details for LOD0 generation, thereby ensuring consistency within the level. Due to the Eq.~\eqref{opti_func} contains three terms, a reasonable assumption is to simulate the function of third term by changing the ratio of weights for the first two terms $f_r$ and $f_s$. In Fig.~\ref{fig:Ablation2}, we compare the results with different weight ratios based on Co-LOD framework. As $\beta$ increases, the main structure of selected architectural model is indeed simplified. However, the removal of essential structural segments resulted in the introduction of random noisy planes to close off the main structure. Overall, co-analysis cannot be simply replaced in accurate LOD generation task. We believe that it is a significant discovery for future research works on architectural reconstruction.

\paragraph{\textbf{Limitations}} Co-LOD can handle curved structures, but its reliance on piecewise planar approximation introduces more errors in curved areas. Tweaking Region Growing~\cite{RegionGrowing} parameters can control the precision of surface approximation and, consequently, the reconstruction error, as demonstrated in Fig.~\ref{curve_limitataion}. A reduced angle threshold enhances accuracy but also results in longer computation times and a higher count of output faces. The Structural Segment Generation algorithm performs well with noisy inputs. However, due to its reliance on spatial discretization for visibility analysis, it overlooks finer structures, as shown in Fig.~\ref{fine_structure}. Even increasing voxel density can help, the computational cost is correspondingly increased. Especially for large-scale scenes, such computational increase for both segment generation and co-analysis is inevitable, as observed in Fig.~\ref{fig:Scalability}. The nonlinear growth stems from the co-analysis stage, mainly due to solving the optimization equation Eq.~\ref{opti_func}. Fortunately, Co-LOD for single building provides an potential solution to address this nonlinear growth issue. The computationally expensive co-analysis can be accomplished through pre-training on large-scale databases, enabling independent control over the LOD generation of each building.

%% file: tables/input.tex
\begin{table}[]
    \caption{Statistics of input scenes, including $\overline{V}$: average vertices, $\overline{F}$ average faces, $\overline{C}$: average connected components, $\overline{I}$: average self-intersections, $\overline{P}$: average detected planes, $\overline{S}$: average segments, $\overline{T}$: average runtime, $M$: model numbers.}
    \label{table:input}
    \resizebox{\columnwidth}{!}{
    \begin{tabular}{l|c|c|c|c|c|c|c|c}
    Scenes                               & $\overline{V}$ & $\overline{F}$ & $\overline{C}$ & $\overline{I}$ & $\overline{P}$ & $\overline{S}$ & $\overline{T}$ & $M$ \\ \hline
    \textit{Metropolis} & 199k           & 271k           & 5.9k           & 363k           & 2476           & 22             & 229.8s         & 54  \\
    \textit{Town}                                 & 148k           & 288k           & 1k             & 15k            & 1378           & 13             & 75.6s          & 27  \\
    \textit{Research Center}                      & 371k           & 543k           & 7k             & 971k           & 3127           & 72             & 776.9s         & 14  \\
    \textit{Composite Scene}                      & 588k           & 867k           & 16k            & 964k           & 3468           & 93             & 987.8s         & 13  \\
    \textit{European City}                        & 59k            & 86k            & 2.3k           & 96k            & 735            & 67             & 522.7s         & 10  \\
    \textit{Campus}                               & 56k            & 70k            & 3.6k           & 125k           & 501            & 22             & 144.8s         & 37  \\
    \textit{Suburbia}                             & 1k             & 2k             & 1              & 3              & 18             & 3              & 56.7s          & 266
    \end{tabular}}
\end{table}

%% file: tables/Comparison_statistics.tex
\begin{table*}[]
  \caption{Statistics and evaluations of different LOD generation methods for related LOD layers. Since PolyFit can only handle buildings from D7 successfully, we separately list quantitative data in D7. Other results are reported based on all scenes. \#V: vertices, \#F: triangles, \#P: polygons, mean: average value of related metric, sd: standard deviation of related metric. For statistic analysis of success rate $r_s$, a failure case means that if the method cannot generate valid output or the running time exceeds one hour. I-LOD and S-LOD present user study outcomes for individual buildings and scenes, respectively. The values represent the percentage each method was selected as the best.}
  \label{table:Comparison}
  \begin{tabular}{c|c|c|c|c|c|cc|cc|c|c}
  \multirow{2}{*}{LOD Layer} & \multirow{2}{*}{Method} & \multirow{2}{*}{\#V} & \multirow{2}{*}{\#F} & \multirow{2}{*}{\#P} & \multirow{2}{*}{${r_s}$} & \multicolumn{2}{c|}{LFD $\downarrow$} & \multicolumn{2}{c|}{HD $\downarrow$} & \multirow{2}{*}{I-LOD $\uparrow$} & \multirow{2}{*}{S-LOD $\uparrow$} \\ \cline{7-10}
                             &                         &                      &                      &                      &                                                  & \multicolumn{1}{c|}{mean}              & sd        & \multicolumn{1}{c|}{mean}              & sd       &                                                &                                                \\ \hline
  \multirow{2}{*}{LOD0 (D7)} & PolyFit                 & 65                   & 51                   & 9                    & 42\%                                             & \multicolumn{1}{c|}{6757.7}            & 2886.0    & \multicolumn{1}{c|}{0.110}             & 0.060    & ——                                             & ——                                             \\
                             & Ours                    & 14                   & 22                   & 9                    & 42\%                                             & \multicolumn{1}{c|}{\textbf{5063.8}}   & 1875.6    & \multicolumn{1}{c|}{\textbf{0.035}}    & 0.025    & ——                                             & ——                                             \\ \hline
  \multirow{4}{*}{LOD0}      & KSR                     & 232                  & 427                  & 124                  & 47.8\%                                           & \multicolumn{1}{c|}{5963.1}            & 4126.3    & \multicolumn{1}{c|}{0.048}             & 0.053    & 10.9\%                                         & 8.7\%                                          \\
                             & QEM                     & 189                  & 125                  & 113                  & 100\%                                            & \multicolumn{1}{c|}{5388.7}            & 2577.4    & \multicolumn{1}{c|}{0.037}             & 0.028    & 3.2\%                                          & 4.7\%                                          \\
                             & RobustLowPoly           & 117                  & 254                  & 250                  & 100\%                                            & \multicolumn{1}{c|}{7146.5}            & 4349.8    & \multicolumn{1}{c|}{0.097}             & 0.069    & 11.7\%                                         & 5.3\%                                          \\
                             & Ours                    & 66                   & 125                  & 40                   & 100\%                                            & \multicolumn{1}{c|}{\textbf{4507}}     & 1785.7    & \multicolumn{1}{c|}{\textbf{0.025}}    & 0.024    & \textbf{77.3\%}                                & \textbf{81.3\%}                                \\ \hline
  \multirow{5}{*}{LOD1}      & KSR                     & 1010                 & 1932                 & 681                  & 98.8\%                                           & \multicolumn{1}{c|}{3793.2}            & 1429.4    & \multicolumn{1}{c|}{0.010}             & 0.011    & 10.9\%                                         & 8.7\%                                          \\
                             & LowPoly                 & 542                  & 1046                 & 645                  & 100\%                                            & \multicolumn{1}{c|}{3899.5}            & 1235.5    & \multicolumn{1}{c|}{0.012}             & 0.006    & ——                                             & ——                                             \\
                             & QEM                     & 5159                 & 3753                 & 3584                 & 100\%                                            & \multicolumn{1}{c|}{3889.2}            & 4075.2    & \multicolumn{1}{c|}{\textbf{0.005}}    & 0.005    & 3.2\%                                          & 4.7\%                                          \\
                             & RobustLowPoly           & 1865                 & 3750                 & 3594                 & 100\%                                            & \multicolumn{1}{c|}{4961.7}            & 3770.9    & \multicolumn{1}{c|}{0.073}             & 0.068    & 11.7\%                                         & 5.3\%                                          \\
                             & Ours                    & 1888                 & 3755                 & 1352                 & 100\%                                            & \multicolumn{1}{c|}{\textbf{3626.5}}   & 1481.2    & \multicolumn{1}{c|}{\textbf{0.005}}    & 0.004    & \textbf{77.3\%}                                & \textbf{81.3\%}                                \\ \hline
  \multirow{4}{*}{LOD2}      & KSR                     & 6648                 & 12898                & 7723                 & 70.0\%                                           & \multicolumn{1}{c|}{3026.3}            & 1103.3    & \multicolumn{1}{c|}{0.007}             & 0.013    & 10.9\%                                         & 8.7\%                                          \\
                             & QEM                     & 53629                & 31501                & 31239                & 100\%                                            & \multicolumn{1}{c|}{9892}              & 12607.1   & \multicolumn{1}{c|}{0.002}             & 0.001    & 3.2\%                                          & 4.7\%                                          \\
                             & RobustLowPoly           & 15719                & 31535                & 29982                & 100\%                                            & \multicolumn{1}{c|}{\textbf{1975.4}}   & 405.7     & \multicolumn{1}{c|}{\textbf{0.001}}    & 0.001    & 11.7\%                                         & 5.3\%                                          \\
                             & Ours                    & 15811                & 31535                & 14516                & 100\%                                            & \multicolumn{1}{c|}{2237.2}            & 432.4     & \multicolumn{1}{c|}{\textbf{0.001}}    & 0.001    & \textbf{77.3\%}                                & \textbf{81.3\%}                               
  \end{tabular}
\end{table*}

%% file: tables/single_anlysis.tex
\begin{table}[]
    \renewcommand\arraystretch{1.1}
    \caption{Statistics of Co-LOD in different LOD generation tasks on \textit{Campus}. S: single building analysis, Co: co-analysis based on the scene, superscript numbers: LOD layer, \#F: triangles, \#V: vertices, $T_s$: segments generation time cost, $T_c$: co-analysis time cost, $T_p$: polygonal mesh extraction time cost.}
    \label{table:signle_analysis}
    \begin{tabular}{c|c|c|c|c|c|c|c}
    Task   & \#F  & \#V  & HD    & LFD  & $T_s$                    & $T_c$                    & $T_p$                   \\ \hline
    $S^0$  & 421  & 188  & 0.023 & 3633 & \multirow{2}{*}{2827.1s} & \multirow{2}{*}{167.3s}  & \multirow{2}{*}{534.6s} \\
    $S^1$  & 6898 & 3510 & 0.001 & 1873 &                          &                          &                         \\ \cline{6-8} 
    $Co^0$ & 346  & 179  & 0.021 & 3656 & \multirow{2}{*}{2834.6s} & \multirow{2}{*}{1972.2s} & \multirow{2}{*}{549.3s} \\
    $Co^1$ & 6775 & 3417 & 0.001 & 1889 &                          &                          &                        
    \end{tabular}
\end{table}

%% file: tables/nr.tex
\begin{table}[]
    \caption{Statistical data for varying $n_r$ values: HD and LFD between the original models and related highest LOD layer models, $T_s$: structural segment generation time cost, $T_p$: polygonal mesh extraction time cost, SDR: calculated using $n_r=200$ as the benchmark.}
    \label{table:nr}
    \begin{tabular}{c|cc|c|c|c|c}
    \multirow{2}{*}{Values of $n_r$} & \multicolumn{2}{c|}{HD}            & \multirow{2}{*}{LFD} & \multirow{2}{*}{$T_s$} & \multirow{2}{*}{$T_p$} & \multirow{2}{*}{SDR} \\ \cline{2-3}
                                                       & \multicolumn{1}{c|}{mean}  & sd    &                      &                        &                        &                      \\ \hline
    5                                                  & \multicolumn{1}{c|}{0.236} & 0.12  & 5305                 & 39.3s                  & 2.7s                   & 2.3\%                \\
    10                                                 & \multicolumn{1}{c|}{0.103} & 0.036 & 3803.2               & 42.5s                  & 2.9s                   & 67.8\%               \\
    30                                                 & \multicolumn{1}{c|}{0.003} & 0.002 & 2499.6               & 49.6s                  & 7.3s                   & 96.3\%               \\
    60                                                 & \multicolumn{1}{c|}{0.002} & 0.002 & 2397.3               & 61.1s                  & 12.5s                  & 97.5\%               \\
    100                                                & \multicolumn{1}{c|}{0.001} & 0.001 & 2375.7               & 78.3s                  & 21.6s                  & 97.3\%               \\
    150                                                & \multicolumn{1}{c|}{0.001} & 0.001 & 2374.9               & 102.7s                 & 43.2s                  & 97.2\%               \\
    200                                                & \multicolumn{1}{c|}{0.001} & 0.001 & 2375.2               & 125.3s                 & 95.3s                  & 100\%               
    \end{tabular}
\end{table}

%% file: src/05-future.tex
\section{Conclusions}
In this paper, we introduce Co-LOD, a novel method for controllable LOD generation with semantic consistency. It signifies pioneering efforts in data-driven LOD generation for urban scenes, aimed at addressing the challenges faced by deep learning solutions in precise reconstruction and LOD control under the current insufficiency of data. The process begins with a robust segmentation method to extract fundamental structural units from architectural models. Subsequently, Co-LOD employs joint structural analysis to facilitate progressive LOD generation. This involves a detailed comparison of segment-based similarities, establishing precise constraints that guarantee semantic consistency across various models. Our experiments validate that Co-LOD uniquely achieves LOD generation with significant semantic consistency. Looking ahead, we plan to expand Co-LOD's capabilities by integrating a generative framework to enhance its completion ability. Additionally, we aim to develop multi-resolution texture mapping for different LOD layers, which will contribute to more realistic rendering of architectural models.